\documentclass[a4paper]{article}

\usepackage{amsmath}
\usepackage{amssymb}
\usepackage{bbm}
\usepackage{graphicx}
\usepackage{comment}
\usepackage{color}
\usepackage{natbib}
\usepackage{calc,enumitem}

\newcommand{\diff}{\mathrm{d}}

\newcommand{\dyn}[1]{\frac{\diff {#1}}{\diff t}}

\newcommand{\avg}[1]{\langle #1 \rangle}

\begin{document}

\textbf{\Large General Relationships between Consumer Dispersal, \\[2pt]
Resource Dispersal and Metacommunity Diversity}

\bigskip \bigskip

{\large Bart Haegeman \& Michel Loreau} \\[12pt]
Centre for Biodiversity Theory and Modelling \\[2pt]
Experimental Ecology Station \\[2pt]
Centre National de la Recherche Scientifique \\[2pt]
Moulis, France \\[6pt]
bart.haegeman@ecoex-moulis.cnrs.fr, michel.loreau@ecoex-moulis.cnrs.fr

\bigskip \bigskip

\textbf{Journal reference} \ B. Haegeman, M. Loreau (2014). General relationships between consumer dispersal, resource dispersal and metacommunity diversity. Ecology Letters 17, 175--184.

\medskip \medskip

%%%%%%%%%%%%%%%%%%%%%%%%%%%%%%%%%%%%%%%%%%%%%%%%%%%%%%%%%%%%%%%%%%%%%%%%%%%%%%%%%%%%%%%
\section*{Abstract}
%%%%%%%%%%%%%%%%%%%%%%%%%%%%%%%%%%%%%%%%%%%%%%%%%%%%%%%%%%%%%%%%%%%%%%%%%%%%%%%%%%%%%%%

One of the central questions of metacommunity theory is how dispersal of organisms affects species diversity.  Here we show that the diversity-dispersal relationship should not be studied in isolation of other abiotic and biotic flows in the metacommunity.  We study a mechanistic metacommunity model in which consumer species compete for an abiotic or biotic resource.  We consider both consumer species specialized to a habitat patch, and generalist species capable of using the resource throughout the metacommunity.  We present analytical results for different limiting values of consumer dispersal and resource dispersal, and complement these results with simulations for intermediate dispersal values.  Our analysis reveals generic patterns for the combined effects of consumer and resource dispersal on the metacommunity diversity of consumer species, and shows that hump-shaped relationships between local diversity and dispersal are not universal.  Diversity-dispersal relationships can also be monotonically increasing or multimodal.  Our work is a new step towards a general theory of metacommunity diversity integrating dispersal at multiple trophic levels.

\newpage

%%%%%%%%%%%%%%%%%%%%%%%%%%%%%%%%%%%%%%%%%%%%%%%%%%%%%%%%%%%%%%%%%%%%%%%%%%%%%%%%%%%%%%%
\section*{Introduction}
%%%%%%%%%%%%%%%%%%%%%%%%%%%%%%%%%%%%%%%%%%%%%%%%%%%%%%%%%%%%%%%%%%%%%%%%%%%%%%%%%%%%%%%

Ecological communities are governed by processes at various spatial scales \citep{MacArthur1967,Ricklefs1987,Levin1992}.  One of the tools to study spatial scales in ecology is the metacommunity concept \citep{Leibold2004,Holyoak2005}.  A metacommunity is a set of communities in a patchy habitat;  communities in different patches are connected by dispersal (or, synonymously, by migration).  Metacommunity models allow us to study the effect of dispersal on the structure and functioning of communities at the local scale, that is, at the scale of each community, and at the regional scale, that is, at the scale of the metacommunity as a whole.

One of the central questions of metacommunity theory is how dispersal affects local and regional diversity.  A standard theoretical argument decomposes the diversity-dispersal relationship into three parts \citep{Loreau2003a,Mouquet2003,Leibold2004}.  First, when dispersal is weak, local communities are essentially isolated.  Local diversity is low due to competitive exclusion at the local scale; regional diversity is high due to spatial heterogeneity between patches.  Second, for moderate dispersal, species disperse from patches where they thrive to patches where they cannot survive without dispersal.  Hence, local diversity increases, while regional diversity remains constant or decreases slowly.  This mixing of local communities continues until local and regional diversity are equal.  Third, when dispersal is strong, the metacommunity is homogenized and competitive exclusion acts at the regional scale.  Both local and regional diversity decrease.  In summary, theory predicts a hump-shaped relationship between local diversity and dispersal and a monotonically decreasing relationship between regional diversity and dispersal.

Numerous experimental studies have measured the diversity-dispersal relationship by manipulating dispersal in microbial, plant and animal metacommunities \citep{Logue2011}.  \citet{Cadotte2006a} conducted a meta-analysis of 50 such experiments.  He found that local diversity increases with dispersal for weak to moderate dispersal, in agreement with metacommunity theory.  However, he obtained ambiguous results for the strong-dispersal part of the diversity-dispersal relationship.  Some studies found that local and regional diversity decrease with increasing dispersal, while other studies suggest that local and regional diversity are unaffected by dispersal when dispersal is strong \citep{Forbes2002,Kneitel2003,Howeth2010,Matthiessen2010}.

Existing theory considers diversity and dispersal of a group of species in isolation of other spatial flows in the metacommunity.  Relaxing this assumption may lead to different predictions, as has been advocated by meta-ecosystem theory \citep{Loreau2003b,Massol2011}.  In particular, it is commonly assumed that a metacommunity is homogenized when dispersal is strong.  But even if the pool of species of which we track diversity is homogeneously distributed, this may not be the case for the entire ecosystem including resources and consumers.  We hypothesize that dispersal at lower or higher trophic levels affects the diversity-dispersal relationship at the focal trophic level.  Here we build a theory that takes into account this extra layer of complexity, focusing on the effect of dispersal on species diversity.

To do so, we introduce a spatial consumer-resource model.  As in existing metacommunity models, we look at a set of interconnected patches in which a resource is consumed locally, and consumers can disperse between patches \citep{Loreau2003a,Loreau2010}.  But contrary to existing metacommunity models, we also consider resource dispersal.  Furthermore, we establish a connection with the theory of resource access limitation \citep{Huston1994,Loreau1998}.  The latter theory predicts that resource dispersal intensifies competition between consumer species.  Thus, in contrast to previous theories, which have dealt with consumer and resource dispersal separately, we investigate the combined effects of consumer and resource dispersal on metacommunity diversity.

More specifically, we model a single limiting resource that is consumed by a pool of consumer species.  The efficiency with which the resource is consumed varies spatially, and differs between species.  We focus on competition for the spatially distributed resource between specialist and generalist consumer species.  Specialists are able to use the resource efficiently in particular patches.  Generalists cannot outcompete specialists in any single local community, but can use the resource throughout the metacommunity.  Their resource use averaged over the patches is more efficient than the average resource use of specialist species.  We study which dispersal conditions, for both consumers and the resource, favor specialists or generalists.

Thus we address three questions in this work: (1) how metacommunity diversity depends on consumer and resource dispersal, (2) what diversity-dispersal relationships are expected for spatial resource competition, and (3) what dispersal values promote specialist or generalist species.  To answer these questions, we first derive analytical results for a number of limiting cases, assuming that consumer and resource dispersal are either very small or very large.  We then use numerical simulations to investigate metacommunities for intermediate dispersal values and to describe their generic diversity-dispersal relationships.

%%%%%%%%%%%%%%%%%%%%%%%%%%%%%%%%%%%%%%%%%%%%%%%%%%%%%%%%%%%%%%%%%%%%%%%%%%%%%%%%%%%%%%%
\section*{Spatial consumer-resource model}
%%%%%%%%%%%%%%%%%%%%%%%%%%%%%%%%%%%%%%%%%%%%%%%%%%%%%%%%%%%%%%%%%%%%%%%%%%%%%%%%%%%%%%%

We present a mechanistic consumer-resource model to explore the effects of dispersal on species diversity.  We consider one limiting resource and several consumer species, all spatially distributed over habitat patches.  We assume, as in previous metacommunity models \citep{Loreau1999,Loreau2003a,Mouquet2003}, that resource consumption rates of consumers differ between patches, \emph{i.e.}, that their growth rates depend on local environmental conditions, such as temperature, acidity, or the presence of a natural enemy.  Competition between consumer species is determined by their patch-dependent resource consumption rates.

We denote by $M$ the number of habitat patches in the metacommunity, and by $S$ the number of consumer species competing for the resource.  The dynamical variables of the model are the biomass of consumer species $i$ in patch $k$, denoted by $N_{ik}$, and the amount of resource in patch $k$, denoted by $R_k$.  The dynamical equations are
\begin{equation}
 \begin{aligned}
 \dyn{N_{ik}} &= e\,c_{ik} R_k N_{ik} - m N_{ik}
  + \alpha \big( \avg{N_{i}} - N_{ik} \big) \\
 \dyn{R_k} &= g_k(R_k) - \sum_i c_{ik} R_k N_{ik}
  + \beta \big( \avg{R} - R_k \big).
 \end{aligned}
\end{equation}
The brackets in $\avg{N_{i}}$ and $\avg{R}$ stand for the average over patches, that is, $\avg{N_{i}} = \frac{1}{M} \sum_k N_{ik}$ and $\avg{R} = \frac{1}{M} \sum_k R_{k}$.

Species $i$ in patch $k$ consumes the resource at rate $c_{ik}$, converts it to new biomass with efficiency $e$ and dies at rate $m$.  For simplicity, and following \citet{Loreau2003a}, we assume that efficiency $e$ and mortality rate $m$ are patch- and species-independent.  The resource in patch $k$ changes at rate $g_k(R_k)$, with
\begin{align*}
 &&&& g_k(R_k) &= a\,(A_k - R_k) && \text{for an abiotic resource,} &&&& \\
 &&&& g_k(R_k) &= b\,R_k (B_k - R_k) && \text{for a biotic resource.} &&&&
\end{align*}
In the case of an abiotic resource, the resource in patch $k$ is supplied at rate $a\,A_k$ and lost at rate $a$.  In the case of a biotic resource, the resource in patch $k$ has intrinsic growth rate $b\,B_k$ and carrying capacity $B_k$.  Parameters $A_k$ and $B_k$ can be interpreted as patch fertilities.  If patch $k$ is empty (no consumers) and isolated (no dispersal), the equilibrium amount of resource is equal to $A_k$ or $B_k$.  We assume that patch fertilities differ between patches.

Consumer species disperse between patches at rate $\alpha$ and the resource disperses (that is, migrates) between patches at rate $\beta$.  As in previous metacommunity models \citep{Loreau1999,Loreau2003a,Mouquet2003}, we model the dispersal process in a minimal way:  dispersal is assumed to be patch-, species- and density-independent.  These simplifying assumptions allow us to focus on the general effects of consumer and resource dispersal.  Consumer dispersal can be much larger than resource dispersal (\emph{e.g.}, plants competing for a soil nutrient such as phosphorus), of the same order of magnitude (\emph{e.g.}, zooplankton species competing for phytoplankton, both undergoing passive dispersal) or much smaller (\emph{e.g.}, bacteria trapped in a biofilm competing for a freely flowing nutrient).  In habitats without specific barriers to dispersal, however, consumers are typically more mobile than resources \citep{McCann2005}.

We are interested in the equilibrium composition of the metacommunity.  It can be shown that in a metacommunity with $M$ patches at most $M$ species can persist (see Appendix~S1).  We investigate how the equilibrium metacommunity composition depends on the model parameters.  In particular, we formulate our results in terms of specialist and generalist consumers.  A consumer species specialized on patch $k$ has large consumption rate $c_{ik}$.  A generalist consumer species has large average consumption rate $\avg{c_i} = \frac{1}{M} \sum_k c_{ik}$.  No species is expected to be specialized on a large number of patches, or to be simultaneously a good specialist and a good generalist.  We assume that the consumer species' consumption rates are subject to specialist-generalist trade-offs \citep{Kneitel2004}.

%%%%%%%%%%%%%%%%%%%%%%%%%%%%%%%%%%%%%%%%%%%%%%%%%%%%%%%%%%%%%%%%%%%%%%%%%%%%%%%%%%%%%%%
\section*{Four limiting cases}
%%%%%%%%%%%%%%%%%%%%%%%%%%%%%%%%%%%%%%%%%%%%%%%%%%%%%%%%%%%%%%%%%%%%%%%%%%%%%%%%%%%%%%%

Model~(1) describes a pool of $S$ consumer species competing for a single limiting resource distributed over $M$ patches.  We are interested in how the equilibrium metacommunity composition depends on consumer dispersal $\alpha$ and resource dispersal $\beta$.  It is difficult (if not impossible) to study model~(1) analytically for arbitrary dispersal values $\alpha$ and $\beta$.  However, it is possible to obtain analytical results by assuming that consumer dispersal and resource dispersal are either very small or very large.  In this section we define and investigate four limiting cases for dispersal values $\alpha$ and $\beta$.  The predictions of the limiting cases are useful to understand the model behavior for arbitrary dispersal values $\alpha$ and $\beta$, as we show in the next section.

The four limiting cases are represented schematically in Figure~1: \\[-16pt]
\begin{itemize}
\item When both $\alpha$ and $\beta$ are small (case I), local communities are isolated.  In each local community $S$ consumer species compete for the resource.  No species persists if patch fertility is too small (see Appendix~S4 for mathematical details).  If patch fertility is sufficiently large, the species that uses the resource most efficiently excludes the other $S-1$ species.  That is, the consumer species that is most specialized on the resource in the patch wins the local competition.  Local diversity is small, but regional diversity is typically large, because the most efficient consumer species differ between patches.  Several specialist species coexist at the regional scale.
%%%
\item When $\alpha$ is large and $\beta$ is small (case II), patches are permeable from the viewpoint of the consumers, but are isolated from the viewpoint of the resource.  Hence, consumers compete regionally for the locally isolated resource.  The resource bound to each of the $M$ local communities corresponds effectively to $M$ distinct resources.  Indeed, in the limit $\alpha\to\infty$ and $\beta=0$ model~(1) reduces to a model of $S$ species competing for $M$ resources (see Appendix~S2),
\begin{equation}
 \begin{aligned}
 \dyn{\avg{N_i}} &= e \sum_k \frac{c_{ik}}{M} R_k \avg{N_i} - m \avg{N_i} \\
 \dyn{R_k} &= g_k(R_k) - \sum_i c_{ik} R_k \avg{N_i}.
 \end{aligned}
\end{equation}
At equilibrium at most $M$ species persist.  The set of persisting species depends on the model parameters, and can be determined by applying non-spatial resource competition theory \citep{Tilman1982,Grover1997}.  Local and regional diversity are equal, and can be small or large depending on the outcome of resource competition.
%%%
\item When $\alpha$ is small and $\beta$ is large (case III), patches are permeable from the viewpoint of the resource, but are isolated from the viewpoint of the consumers.  Hence, locally isolated consumers compete regionally for the resource.  A consumer species bound to each of the $M$ local communities corresponds effectively to $M$ distinct consumer populations.  Hence, there are $M S$ effective consumer populations in total.  Indeed, in this limit model~(1) reduces to a model of $M S$ species competing for one resource (see Appendix~S2),
\begin{equation}
 \begin{aligned}
 \dyn{N_{ik}} &= e\,c_{ik} \avg{R} N_{ik} - m N_{ik} \\
 \dyn{\avg{R}} &= G(\avg{R}) - \sum_{i,k} \frac{c_{ik}}{M} \avg{R} N_{ik}.
 \end{aligned}
\end{equation}
with
\begin{align*}
 &&&& G(\avg{R}) &= a\,(\avg{A} - \avg{R}) && \text{for an abiotic resource,} &&&& \\
 &&&& G(\avg{R}) &= b\,\avg{R}\,(\avg{B} - \avg{R}) && \text{for a biotic resource.} &&&&
\end{align*}
\hspace{-3.5pt}No species persists if patch fertility is too small (see Appendix~S4).  If patch fertility is sufficiently large, the consumer population that uses the resource most efficiently excludes all the other populations.  That is, at equilibrium only a single patch is occupied, that specific patch is occupied by a single species, and that specific species consumes the regionally homogenized resource.  The consumer species that is most specialized on a local resource dominates the entire metacommunity.  Local and regional diversity are equal and small.
%%%
\item When both $\alpha$ and $\beta$ are large (case IV), the metacommunity is homogenized both from the viewpoint of the consumers and from that of the resource.  The spatial structure of the metacommunity dissolves; the $S$ consumer species compete for the resource at the regional scale.  The reduced model is (see Appendix~S2),
\begin{equation}
 \begin{aligned}
 \dyn{\avg{N_i}} &= e \avg{c_i} \avg{R} \avg{N_i} - m \avg{N_i} \\
 \dyn{\avg{R}} &= G(\avg{R}) - \sum_i \avg{c_i} \avg{R} \avg{N_i}.
 \end{aligned}
\end{equation}
No species persists if patch fertility is too small (see Appendix~S4).  If patch fertility is sufficiently large, the species that uses the resource most efficiently averaged over spatial heterogeneity excludes the other species.  That is, the most efficient generalist species dominates the metacommunity.  Local and regional diversity are equal and small.
\end{itemize}

%%%%% FIGURE 1 %%%%%
\begin{figure}
\begin{center}
\includegraphics[width=.42\textwidth]{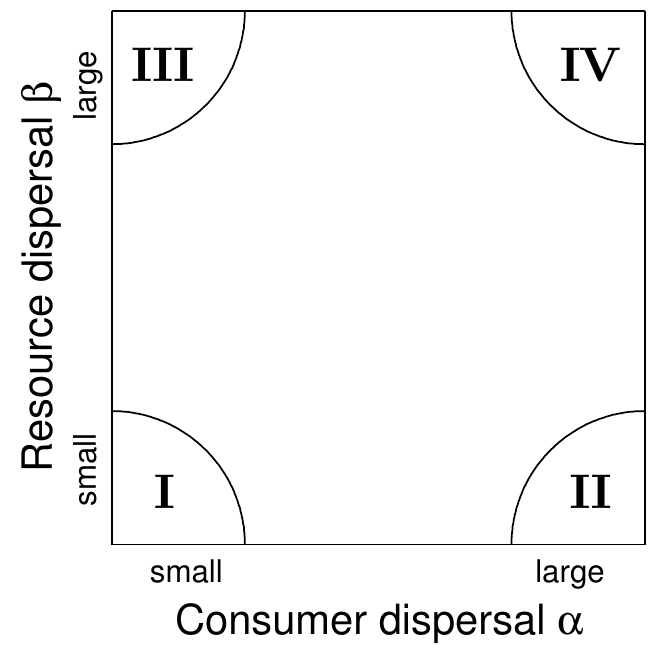}
\hspace{.01\textwidth}
\parbox[b]{.54\textwidth}{\linespread{1} \selectfont
\begin{enumerate}[label=\bf\Roman{enumi}]
%\begin{description}[labelwidth=\widthof{\bf III}]
\item Isolated local communities \\ 
 $\rightarrow$ Each local community is dominated \\
 \mbox{\textcolor{white}{$\rightarrow$}} by best specialist consumer
\item Regional consumers, $M$ local resources \\
 $\rightarrow$ At most $M$ consumers coexist regionally \\
 \mbox{\textcolor{white}{$\rightarrow$}} (specialists and/or generalists)
\item Local consumers, one regional resource \\
 $\rightarrow$ One patch is dominated by the overall \\
 \mbox{\textcolor{white}{$\rightarrow$}} best specialist consumer; other patches \\
 \mbox{\textcolor{white}{$\rightarrow$}} are (almost) empty of consumers
\item Homogeneous metacommunity \\
 $\rightarrow$ The metacommunity is dominated \\
 \mbox{\textcolor{white}{$\rightarrow$}} by best generalist consumer \\[-24pt]
\end{enumerate}}
%\end{description}
\caption{Four limiting cases of consumer and resource dispersal, and the resulting metacommunity structure.  Consumer dispersal $\alpha$ and resource dispersal $\beta$ affect the competition between consumer species for a spatially distributed resource.  We analyze four limiting cases: (I) both $\alpha$ and $\beta$ small; (II) $\alpha$ large and $\beta$ small; (III) $\alpha$ small and $\beta$ large; (IV) both $\alpha$ and $\beta$ large.  These limiting cases provide four reference points for the relationship between dispersal and metacommunity diversity.  They are extended to intermediate dispersal values in Figures~2 and 3.}
\end{center}
\end{figure}

The analysis of these limiting cases provides four reference points for the relationship between consumer dispersal, resource dispersal and metacommunity diversity.  When both consumer dispersal $\alpha$ and resource dispersal $\beta$ are small, each patch is dominated by a single consumer species and the dominant species differ between patches.  When consumer dispersal $\alpha$ is large (and $\beta$ small), the patch compositions mix and the regional competition for the locally isolated resource can have different outcomes.  When resource dispersal $\beta$ is large (and $\alpha$ small), the species that is most specialized on its patch excludes the other species.  When both consumer dispersal $\alpha$ and resource dispersal $\beta$ are large, the most efficient consumer species averaged over spatial heterogeneity, that is, the best generalist species, excludes the other species.

%%%%%%%%%%%%%%%%%%%%%%%%%%%%%%%%%%%%%%%%%%%%%%%%%%%%%%%%%%%%%%%%%%%%%%%%%%%%%%%%%%%%%%%
\section*{Diversity-dispersal relationships}
%%%%%%%%%%%%%%%%%%%%%%%%%%%%%%%%%%%%%%%%%%%%%%%%%%%%%%%%%%%%%%%%%%%%%%%%%%%%%%%%%%%%%%%

In the previous section we have established some reference points for the relationship between consumer dispersal, resource dispersal and metacommunity composition.  Here we present numerical simulations of model~(1) to extend the previous results to intermediate dispersal values.  First, we study a metacommunity with two patches.  Then, we show that larger metacommunities exhibit similar patterns.  Finally, we connect our results with the experimentally often measured diversity-dispersal relationship.

To perform numerical simulations, we integrated model~(1) numerically over a long time span using the MATLAB solver ode15s.  At the end of each simulation we checked that an equilibrium was reached (by evaluating the right-hand side of equations~(1)) and that the equilibrium was stable (by computing the eigenvalues of the Jacobian).  The simulations suggest that there is a unique stable equilibrium for all parameter values considered in this study.

First, we consider a metacommunity with two patches occupied by two specialist species S1 and S2.  The resource is assumed to be biotic.  The effects of consumer and resource dispersal on equilibrium metacommunity composition and on local and regional diversity are shown in Figures~2 and 3, rows (a--b).  We quantify metacommunity diversity using Shannon diversity, which is more convenient for our purpose than species richness (Appendix~S3 and Figure~S1).  As predicted by the previous section, both species coexist regionally for small $\alpha$ and small $\beta$, the best specialist species (here species S1) dominates for small $\alpha$ and large $\beta$, the best generalist species (here species S1) dominates for large $\alpha$ and large $\beta$, and different scenarios are possible for large $\alpha$ and small $\beta$.  In Figures~2a and 3a, patch fertilities are sufficiently large for both species to persist.  In Figures~2b and 3b, patch fertility $B_1$ is too small to maintain specialist species S1.  The corresponding diversity patterns are similar except for large $\alpha$ and small~$\beta$ (Figure 3, rows (a--b)).

We then add a generalist species G to the two-species two-patch metacommunity (Figures~2 and 3, rows (c--d)).  The generalist species has no effect on the metacommunity composition for small $\alpha$.  For large $\alpha$ and small $\beta$, the metacommunity can have different compositions depending on the patch fertilities.  In Figures~2c and 3c, species G excludes the specialist species.  In Figures~2d and 3d, species S2 and G coexist locally.  For large $\alpha$ and large $\beta$, the generalist species G dominates (if patch fertility is sufficiently large).  The odds for generalist species G to be present in the metacommunity at equilibrium increase when increasing $\alpha$, especially for large $\beta$.  Again, the diversity patterns are similar except for large~$\alpha$ and small~$\beta$ (Figure~3, rows (c--d)).

%%%%% FIGURE 2 %%%%%
\begin{figure}
\includegraphics[width=.96\textwidth]{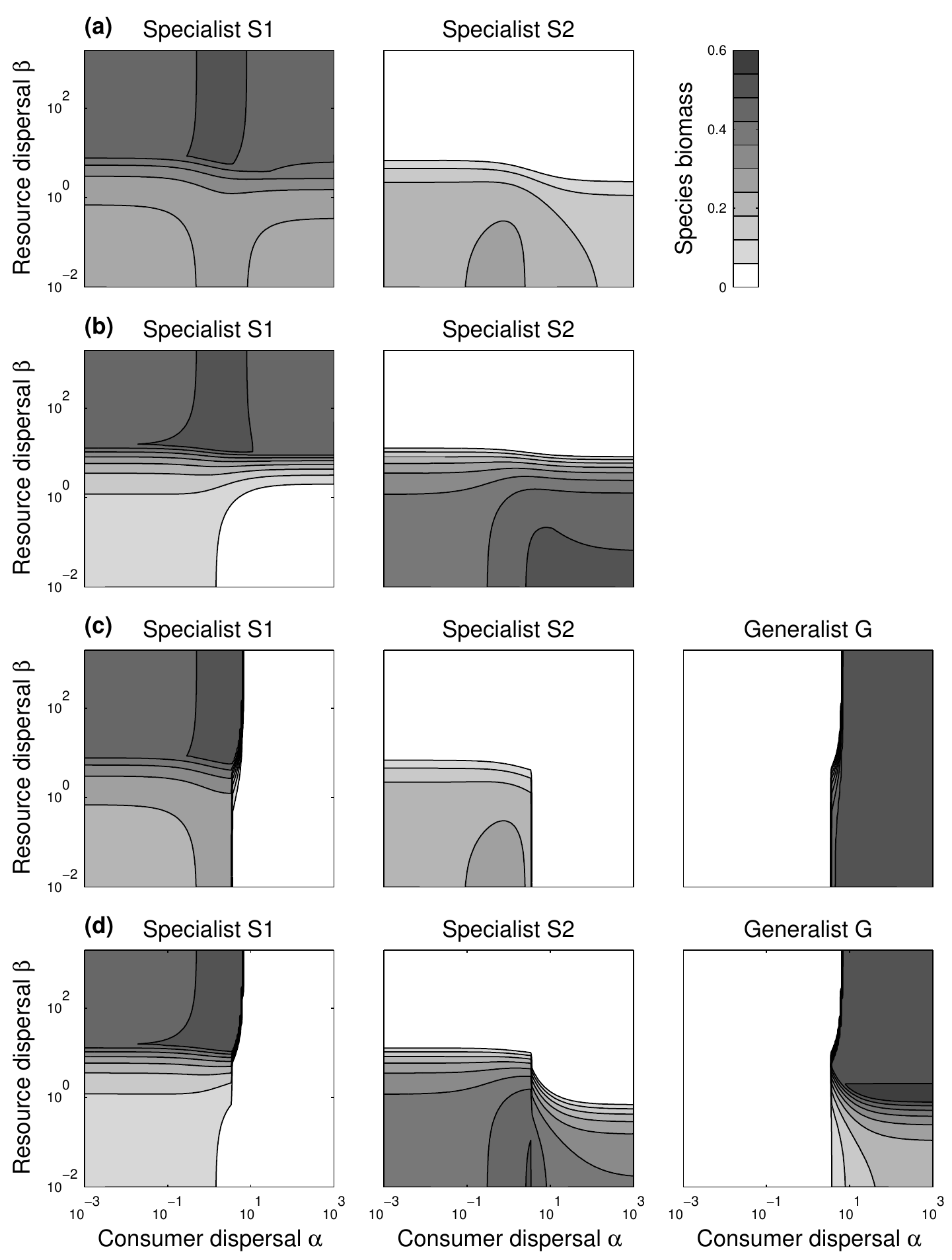}
\caption{Effects of consumer and resource dispersal on the composition of two-patch metacommunities with a biotic resource.  Equilibrium consumer biomass is plotted for four metacommunities.  Rows (a--b): metacommunities with two specialist consumer species S1 and S2;  patch fertilities differ between rows (a) and (b).  Rows (c--d): metacommunities with three consumer species: two specialists S1 and S2 and one generalist G;  patch fertilities differ between rows (c) and (d).  Parameter values:  $e = m = b = 1$.  $c_{11}=3.0$, $c_{12}=0$ for species S1;  $c_{21}=0$, $c_{22}=2.6$ for species S2;  $c_{31}=1.8$, $c_{32}=1.6$ for species G.  (a) $B_1=B_2=1.0$;  (b) $B_1=0.6$, $B_2=1.4$;  (c) $B_1=1.0$, $B_2=1.0$;  (d) $B_1=0.6$, $B_2=1.4$.}
\end{figure}

%%%%% FIGURE 3 %%%%%
\begin{figure}
\begin{center}
\includegraphics[width=.76\textwidth]{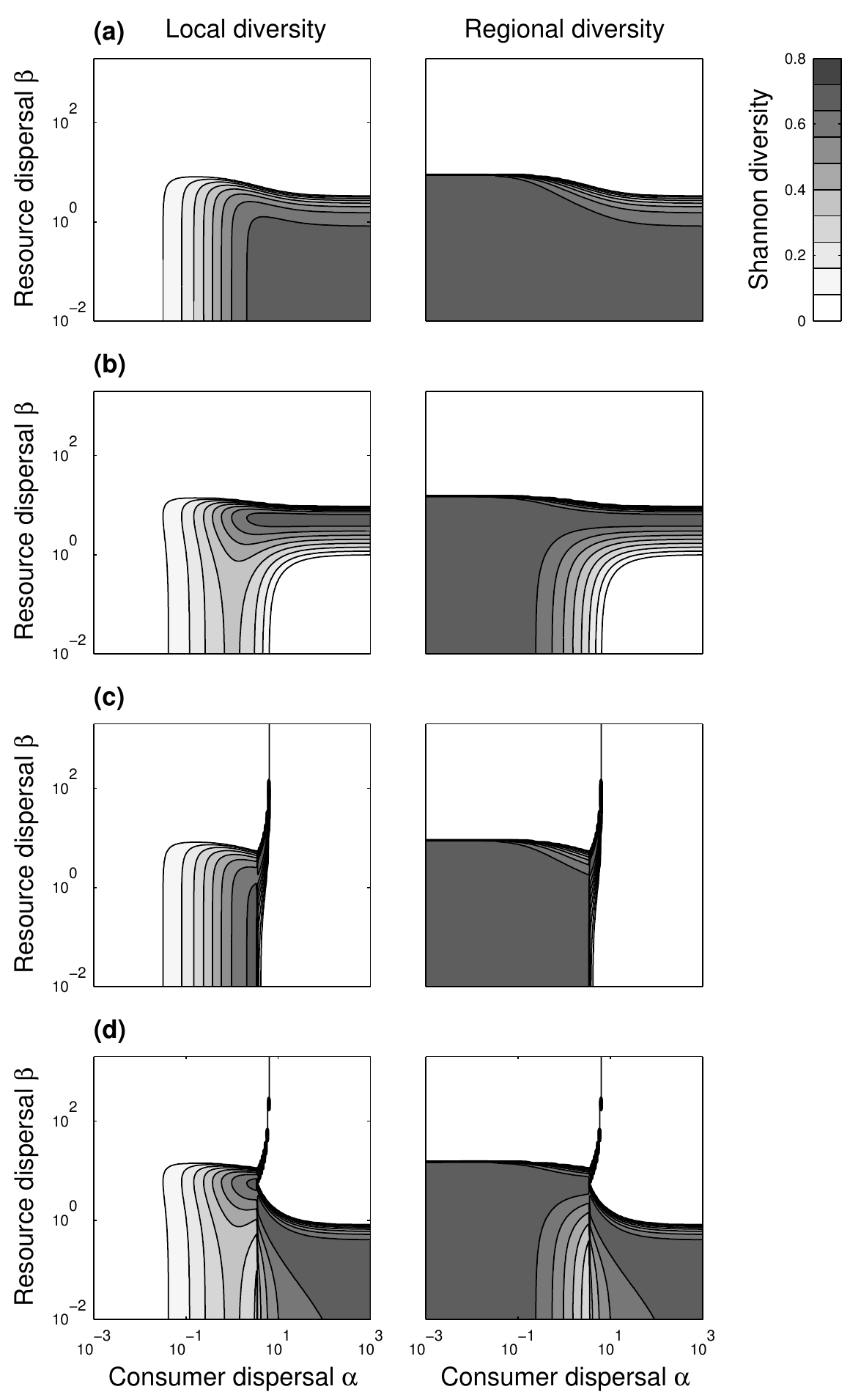}
\caption{Effects of consumer and resource dispersal on the diversity in two-patch metacommunities with a biotic resource.  For the same metacommunities as in Figure~2 we plot local and regional diversity, measured by Shannon diversity (see Appendix~S3).  The first two metacommunities (rows (a--b)) consist of two specialist species S1 and S2.  The last two metacommunities (rows (c--d)) consist of two specialist species S1 and S2 and one generalist species G.}
\end{center}
\end{figure}

Next, we investigate the robustness of the above results with respect to the model assumptions.  First, we study the dependence of metacommunity composition on patch fertilities (Figure~S2, panel (a) and Figure~S3 for the two-species metacommunities;  Figure~S2, panel (b) and Figure~S4 for the three-species metacommunities).  The regions of dispersal values $\alpha$ and $\beta$ for which species can persist, change in accordance with the mechanisms explained above.  Second, we study the effect of replacing a biotic resource by an abiotic resource (compare Figure~S5 with Figure~2 and Figure~S6 with Figure~3).  Taking the same patch fertilities for biotic and abiotic resource, the metacommunity patterns are qualitatively similar.

Third, we investigate whether the results for metacommunities with two patches extend to larger metacommunities.  We used different procedures to generate simulation parameters (Appendix~S5).  A first procedure does not impose a trade-off on the set of consumption rates $c_{ik}$ for species $i$.  A second procedure assumes a linear trade-off, that is, the sum $\sum_k c_{ik}$ is the same for all species $i$.  A third procedure assumes a quadratic trade-off, that is, the sum $\sum_k c_{ik}^2$ is the same for all species $i$.  As explained in Appendix~S5, only the last procedure implements a specialist-generalist trade-off \citep{Kneitel2004}.  It prevents a species from being specialized on a large number of patches, or from being simultaneously a specialist and a generalist.  Nevertheless, we find that the three procedures lead to comparable diversity patterns (Figure~S8), indicating that the patterns we obtained are generic.

Using the parameter generation procedure with a quadratic trade-off, we simulated a large number of metacommunities with $M=5$ patches and $S=20$ species.  Examples of results for four such metacommunities are shown in Figure~S10.  The metacommunity diversity patterns are similar over a large region of dispersal values.  There are qualitative differences between simulations only for large $\alpha$ and small $\beta$;  the outcomes then range from competitive exclusion to the local coexistence of five species (the maximal number of species that can coexist regionally in a metacommunity with five patches), as predicted by our analysis of the corresponding limiting case in the previous section.  The patterns describing the presence of specialist \emph{vs.} generalist consumer species are also similar.  Specialists are favored for small $\alpha$, especially when $\beta$ is large;  generalists are favored for large $\alpha$, especially when $\beta$ is large.  In short, both the diversity patterns and the specialist \emph{vs.} generalist patterns are similar to those for two-patch metacommunities (Figures~2 and 3).

Finally, we investigate which diversity-dispersal relationships are predicted by our model.  There are several ways to take a one-dimensional cross-section of a two-dimensional metacommunity diversity pattern (the two dimensions correspond to consumer dispersal $\alpha$ and resource dispersal $\beta$).  As an illustration, we construct three diversity-dispersal relationships for a metacommunity with $M=5$ patches and $S=20$ species (Figure~4).  For the first relationship, we increase consumer dispersal while keeping resource dispersal small (panel (c)).  Local diversity shows an overall increasing trend despite irregularities;  regional diversity decreases towards local diversity until the two diversities coincide.  For the second relationship, we increase consumer dispersal and resource dispersal simultaneously (panel (d)).  Local diversity shows a hump-shaped pattern;  regional diversity decreases towards local diversity.  For the third relationship, we increase resource dispersal while keeping consumer dispersal small (panel (e)).  Local diversity is small for all dispersal values;  regional diversity decreases steeply from maximal diversity to zero diversity.

%%%%% FIGURE 4 %%%%%
\begin{figure}
\begin{center}
\includegraphics[width=.82\textwidth]{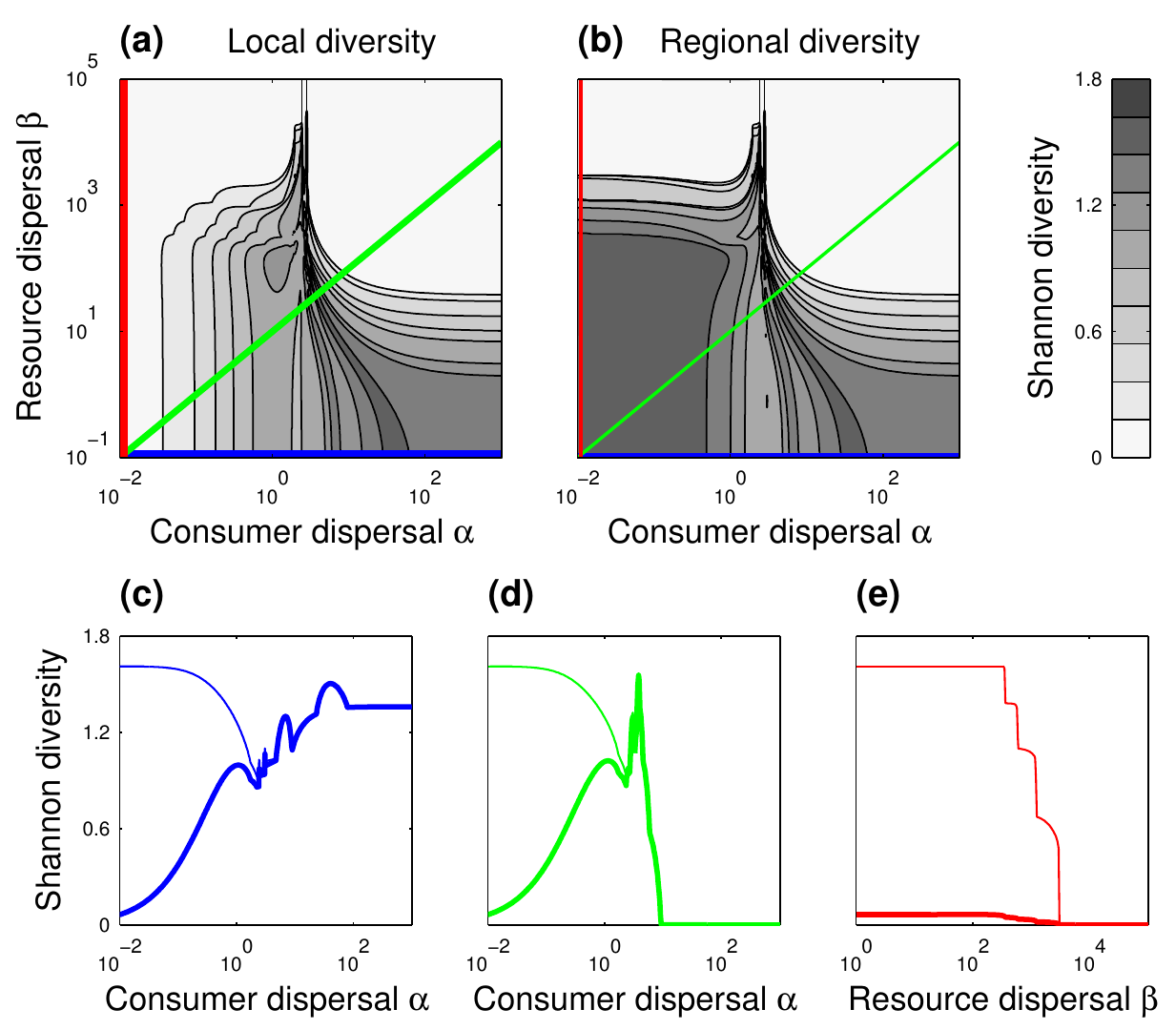}
\end{center}
\caption{Our model predicts a range of one-dimensional diversity-dispersal relationships.  Here we study a metacommunity with $M=5$ patches and $S=20$ species, in which parameters were generated with the procedure described in Appendix~S5.  Local and regional diversity patterns (panels (a--b)) are qualitatively similar to the patterns for two-patch metacommunities (compare with Figure~3).  From the two-dimensional patterns we derive three one-dimensional diversity-dispersal relationships.  Blue cross-section, panel (c): consumer dispersal $\alpha$ varies while keeping resource dispersal $\beta$ small.  Green cross-section, panel (d): consumer dispersal $\alpha$ and resource dispersal $\beta$ vary simultaneously.  Red cross-section, panel (e): resource dispersal $\beta$ varies while keeping consumer dispersal $\alpha$ small.  Thick line: local diversity;  thin line: regional  diversity.}
\end{figure}

We checked the robustness of these findings for a large number of metacommunities using the parameter generation procedures without and with trade-offs (Figures~S9 and S11).  We obtain qualitatively similar diversity-dispersal relationships to those for the example in Figure~4, except for the last part of the first relationship, when consumer dispersal is strong and further increases while resource dispersal is weak and constant.  This part can be slowly increasing, slowly decreasing or steeply decreasing to zero.  This observation is consistent with our results for the limiting case of large $\alpha$ and small $\beta$, for which different outcomes are possible.  In short, our model predicts a wide range of diversity-dispersal relationships.

%%%%%%%%%%%%%%%%%%%%%%%%%%%%%%%%%%%%%%%%%%%%%%%%%%%%%%%%%%%%%%%%%%%%%%%%%%%%%%%%%%%%%%%
\section*{Discussion}
%%%%%%%%%%%%%%%%%%%%%%%%%%%%%%%%%%%%%%%%%%%%%%%%%%%%%%%%%%%%%%%%%%%%%%%%%%%%%%%%%%%%%%%

Our metacommunity model generalizes previous theories by considering patches that are coupled by both consumer dispersal and resource dispersal.  We have shown that both consumer dispersal and resource dispersal strongly affect metacommunity diversity.  When considering the effect of resource dispersal, we recover the predictions of models describing limited resource access \citep{Huston1994,Loreau1998}.  Without dispersal patches are dominated by different consumer species, so that locally dominant consumer species coexist at the regional scale.  Increasing resource dispersal homogenizes the spatial distribution of the resource and increases resource competition between consumer species, even if competing species occupy different patches.  As a result, regional diversity decreases with increasing resource dispersal.

When considering the effect of consumer dispersal, we recover some predictions of previous metacommunity models \citep{Loreau1999,Mouquet2003}.  Local communities, which are dissimilar without dispersal, are mixed when increasing consumer dispersal.  As a result, consumer species can be maintained in patches in which they cannot persist without dispersal (that is, source-sink effects).  The increase in local diversity with consumer dispersal continues until local and regional diversity are equal.  When further increasing consumer dispersal, however, different scenarios are possible.  It is generally argued that large dispersal homogenizes the metacommunity.  As the spatial structure no longer provides a mechanism for regional species coexistence, metacommunity diversity should collapse.  However, this scenario implicitly assumes that increased consumer dispersal also leads to increased resource dispersal.  If this is not the case, that is, if consumer dispersal increases while resource dispersal remains small, consumer species compete for a resource that is isolated in different habitat patches.  We have shown that this situation is equivalent to consumer species competing for distinct ``effective'' resources, that is, resources bound to different patches \citep{Abrams1988}.  Therefore, several (up to the number of patches) consumer species can coexist locally with large consumer dispersal.  Whether this coexistence is realized depends on the outcome of resource competition between consumer species.  For the case of a two-patch metacommunity, the resulting composition can be determined by a graphical analysis, analogous to non-spatial competition for two resources \citep{Tilman1982,Grover1997}.

Thus, our study extends metacommunity theory and unifies it with limited resource access theory.  This unification also provides a broader perspective on the relationship between local diversity and consumer dispersal, which is one of the main patterns predicted by metacommunity theory.  Previous models generically predict hump-shaped relationships \citep{Loreau2003a,Mouquet2003}.  By contrast, our model generically predicts a range of possible relationships.  In particular, if consumer dispersal varies but resource dispersal stays constant, the diversity-dispersal relationship changes from hump-shaped to overall increasing (with irregularities, however), depending on the level of metacommunity diversity for large consumer dispersal and small resource dispersal.  Hence, hump-shaped diversity-dispersal relationships cannot be expected to hold universally.  However, our model does predict hump-shaped relationships if increasing consumer dispersal entails a concomitant increase in resource dispersal.  \citet{Mouquet2003} considered also a non-generic set of parameters for which all species have exactly the same competitive ability at the regional scale, leading to a monotonically increasing relationship.

Our model includes previous predictions of metacommunity theory for the diversity-dispersal relationship as special cases.  Loreau \emph{et al.}'s (2003a) model is closely related to ours, except that it also includes environmental fluctuations.  However, these fluctuations have a small effect on the predicted diversity-dispersal relationship (Appendix~S6).  \citet{Loreau2003a} did not consider resource dispersal and they studied only a specific set of parameter values for which one generalist excludes the other species for large consumer dispersal (and without resource dispersal).  This explains why they predicted a hump-shaped diversity-dispersal relationship.  However, as we have stressed above, this prediction is contingent on a particular choice of parameter values.  The same model can lead to other diversity-dispersal relationships for slightly different parameter values.  For example, by decreasing niche overlap between species, species coexistence is facilitated and the diversity-dispersal relationship reaches higher diversity values for large consumer dispersal (Appendix~S6).  Mouquet \& Loreau's (2003) model is rather different from ours, because it is based on a lottery competition instead of the mechanistic resource competition described by ours.  Nevertheless, it can be interpreted in our framework by noting that in their model the limit of large dispersal homogenizes the metacommunity.  This corresponds to the joint limit of large consumer dispersal and large resource dispersal in our model.  As a result, our theory predicts a hump-shaped diversity-dispersal relationship, as reported by \citet{Mouquet2003}.  Interestingly, although both \citet{Loreau2003a} and \citet{Mouquet2003} predicted that the diversity-dispersal relationship drops to zero at large dispersal, they did so for different reasons.  In \citet{Mouquet2003}, large dispersal homogenizes the metacommunity, so that no diversity can be maintained.  In \citet{Loreau2003a}, large (consumer) dispersal does not homogenize the metacommunity (the resource distribution is heterogeneous), but it increases the competitive advantage of a generalist species, which excludes all other species.  Thus, our theory unifies previous results by considering the combined effects of consumer dispersal and resource dispersal.

It is worth noting, however, that models with a more implicit description of species competition do not necessarily predict hump-shaped diversity-dispersal relationships either.  For example, the metacommunity model with local Lotka-Volterra competitive interactions \citep[used, \emph{e.g.}, in][]{Levin1974,Amarasekare2001},
\begin{equation}
 \dyn{N_{ik}} = \frac{r_{ik} N_{ik}}{K_{ik}}
 \bigg( K_{ik} - \sum_j a_{ijk} N_{jk} \bigg)
 + \alpha \big( \avg{N_i} - N_{ik} \big),
\end{equation}
also leads to a range of diversity-dispersal relationships depending on the choice of parameter values (intrinsic growth rates $r_{ik}$, carrying capacities $K_{ik}$ and competition coefficients $a_{ijk}$).  The predicted relationships are similar to those of our model without resource dispersal.  In fact, there is a formal equivalence between, on the one hand, model~(5) and, on the other hand, model~(1) without resource dispersal and with fast resource dynamics.  This equivalence is analogous to that between the non-spatial Lotka-Volterra competition model and the non-spatial consumer-resource model with fast resource dynamics \citep{MacArthur1972,Abrams2008}.  Hence, model~(5) with large dispersal corresponds to model~(1) with large consumer dispersal and small resource dispersal and not to model~(1) with large consumer dispersal and large resource dispersal.  In other words, model~(5) implicitly assumes a heterogeneous resource distribution, even though consumer dispersal $\alpha$ is large.  This illustrates the relevance of our theory for a larger class of metacommunity models and, more generally, the importance of taking into account spatial resource flows in metacommunity models.

Our results indicate that the experimental setup used to manipulate dispersal can change the diversity-dispersal relationship qualitatively (compare Figure~4, panels (c) and (d)).  If consumer dispersal is varied without affecting resource dispersal (\emph{e.g.}, by sowing different amounts of seeds in plant metacommunities), then a range of diversity-dispersal relationships is possible.  If a variation in consumer dispersal entails a simultaneous variation in resource dispersal (\emph{e.g.}, by transferring different volumes of water in aquatic metacommunities), then the diversity-dispersal relationship should be hump-shaped.  On the whole, the hump-shaped relationship should not be considered as a hallmark of metacommunity structure.  The meta-analysis of \citet{Cadotte2006a} was explicitly directed at detecting a hump-shaped pattern in experimental studies.  It might be more insightful to allow for a wider range of possible relationships, thereby taking into account the experimental setup used to manipulate dispersal.  \citet{Cadotte2006a} obtained ambiguous results for the strong-dispersal part of the diversity-dispersal relationship, precisely where our model predictions deviate from previous metacommunity models.

Our study implies that one-dimensional diversity-dispersal relationships are not as strong an experimental test of metacommunity theory as previously thought.  Two-dimensional relationships, in which consumer dispersal and resource dispersal are varied independently, would be more instructive about the underlying metacommunity processes.  Such a relationship could be measured, for example, in the Metatron, a large-scale experimental setup for multitrophic terrestrial metacommunities \citep{Legrand2012}.  Strong tests of the predicted metacommunity patterns would result from considering a large number of combinations of consumer dispersal and resource dispersal, spanning the range from small to large values (Figure~1).  A preliminary two-dimensional relationship was measured by \citet{Limberger2011}, studying the effects of prey and predator dispersal on prey diversity (rather than predator diversity as in this study).  Several studies have investigated the effect of resource levels on metacommunity structure \citep{Kneitel2003,Cadotte2006b,Matthiessen2010}.  Our model could also serve as a theoretical framework for these experiments (see Figures~S3 and S4).  Finally, we have described the effects of consumer and resource dispersal on the presence of specialist \emph{vs.} generalist species.   We found a simple pattern that to our knowledge has not been described previously \citep{Kneitel2004}.  This pattern can be studied experimentally, or could be useful to interpret observational data \citep{Pandit2009}.

A current challenge in metacommunity theory is to integrate trophic structure into spatial community models \citep{Holt2002,McCann2005,Amarasekare2008a,Amarasekare2008b,Pillai2011}.  Our spatial consumer-resource model, which may be viewed as a metacommunity model with two trophic levels, is a new step towards this goal.  Our analysis of this model was based on an analytical study of limiting cases assuming very small or very large dispersal values, complemented with numerical simulations for intermediate dispersal values.  By combining these tools, we have obtained a broad and detailed understanding of the model.  We suggest that a similar approach will be useful to investigate metacommunity models with more intricate trophic structure.

%%%%%%%%%%%%%%%%%%%%%%%%%%%%%%%%%%%%%%%%%%%%%%%%%%%%%%%%%%%%%%%%%%%%%%%%%%%%%%%%%%%%%%%
\section*{Acknowledgements}
%%%%%%%%%%%%%%%%%%%%%%%%%%%%%%%%%%%%%%%%%%%%%%%%%%%%%%%%%%%%%%%%%%%%%%%%%%%%%%%%%%%%%%%

This work was supported by the TULIP Laboratory of Excellence (ANR-10-LABX-41).

%%%%%%%%%%%%%%%%%%%%%%%%%%%%%%%%%%%%%%%%%%%%%%%%%%%%%%%%%%%%%%%%%%%%%%%%%%%%%%%%%%%%%%%
% References
%%%%%%%%%%%%%%%%%%%%%%%%%%%%%%%%%%%%%%%%%%%%%%%%%%%%%%%%%%%%%%%%%%%%%%%%%%%%%%%%%%%%%%%

\newpage

\renewcommand{\thesection}{S\arabic{section}}
\setcounter{section}{0}
\renewcommand{\theequation}{S\arabic{equation}}
\setcounter{equation}{0}
\renewcommand{\thefigure}{S\arabic{figure}}
\setcounter{figure}{0}
\renewcommand{\thetable}{S\arabic{table}}
\setcounter{table}{0}

\begin{center}
\textsc{\LARGE Supplementary Information}
\end{center}

\bigskip

\noindent {\bf Supplementary Text}
\begin{enumerate}
\item[] Appendix S1 \ Number of coexisting species
\item[] Appendix S2 \ Model reduction for large dispersal
\item[] Appendix S3 \ Quantifying metacommunity diversity
\item[] Appendix S4 \ Analysis of four limiting cases
\item[] Appendix S5 \ Generating simulation parameters
\item[] Appendix S6 \ Comparison with Loreau et al.~(2003a)
\end{enumerate}

\bigskip

%\noindent {\bf Supplementary Tables}
%\begin{enumerate}
%\item[] Table S1 \ 
%\item[] Table S2 \ 
%\end{enumerate}

\noindent {\bf Supplementary Figures}
\begin{enumerate}
\item[] Figure S1 \ \begin{minipage}[t]{11.4cm} Quantifying metacommunity diversity using species richness and Shannon diversity \end{minipage}
\item[] Figure S2 \ \begin{minipage}[t]{11.4cm} Effects of patch fertility on two-patch metacommunities for limiting case $\alpha\to\infty$ and $\beta=0$ \end{minipage}
\item[] Figure S3 \ \begin{minipage}[t]{11.4cm} Effects of dispersal and patch fertility on two-patch two-species metacommunity \end{minipage}
\item[] Figure S4 \ \begin{minipage}[t]{11.4cm} Effects of dispersal and patch fertility on two-patch three-species metacommunity \end{minipage}
\item[] Figure S5 \ \begin{minipage}[t]{11.4cm} Effects of consumer and resource dispersal on composition of two-patch metacommunity with abiotic resource \end{minipage}
\item[] Figure S6 \ \begin{minipage}[t]{11.4cm} Effects of consumer and resource dispersal on diversity of two-patch metacommunity with abiotic resource \end{minipage}
\item[] Figure S7 \ \begin{minipage}[t]{11.4cm} Randomly generating species consumption rates \end{minipage}
\item[] Figure S8 \ \begin{minipage}[t]{11.4cm} Diversity patterns for five-patch metacommunities without and with trade-offs \end{minipage}
\item[] Figure S9 \ \begin{minipage}[t]{11.4cm} One-dimensional diversity-dispersal relationships for five-patch metacommunities without and with trade-offs \end{minipage}
\item[] Figure S10 \ \begin{minipage}[t]{11.4cm} Diversity patterns for five-patch metacommunities with quadratic trade-off \end{minipage}
\item[] Figure S11 \ \begin{minipage}[t]{11.4cm} One-dimensional diversity-dispersal relationships for five-patch metacommunities with quadratic trade-off \end{minipage}
\item[] Figure S12 \ \begin{minipage}[t]{11.4cm} Diversity-dispersal relationships predicted by Loreau et al.~(2003a) \end{minipage}
\end{enumerate}

\newpage

%%%%%%%%%%%%%%%%%%%%%%%%%%%%%%%%%%%%%%%%%%%%%%%%%%%%%%%%%%%%%%%%%%%%%%%%%%%%%%%%%%%%%%%
\section*{Appendix S1 \ Number of coexisting species}
%%%%%%%%%%%%%%%%%%%%%%%%%%%%%%%%%%%%%%%%%%%%%%%%%%%%%%%%%%%%%%%%%%%%%%%%%%%%%%%%%%%%%%%
\renewcommand{\theequation}{S\arabic{equation}}
\setcounter{equation}{0}
\renewcommand{\thefigure}{S\arabic{figure}}
\setcounter{figure}{0}

Here we study the equilibrium conditions of model~(1), and in particular, the number of coexisting species at the regional scale.  At equilibrium the conditions $\dyn{N_{ik}} = 0$ and $\dyn{R_k} = 0$ have to be satisfied simultaneously.  We analyze the conditions $\dyn{N_{ik}} = 0$ for a specific species $i$ and for all patches $k=1,\ldots,M$.  To do so, we consider the amounts of resource $R_k$ to be fixed.  Thus, we have $M$ equations for the $M$ species biomasses $N_{ik}$ with $k=1,\ldots,M$,
\begin{equation}
 e\,c_{ik} R_k N_{ik} - m N_{ik} + \alpha \sum_\ell N_{i\ell} - \alpha M N_{ik} = 0.
 \label{eq:howmany1}
\end{equation}
These equations are linear in $N_{ik}$.  Hence, $N_{ik} = 0$ for $k=1,\ldots,M$ is a solution.  Clearly, if this would be the only solution, species $i$ would be absent from the metacommunity at equilibrium.  We look for other, non-trivial solutions.  This is possible only if the coefficient matrix of (\ref{eq:howmany1}) is singular, that is, if
\begin{equation}
 \det \left( \begin{bmatrix}
 d_1 & \alpha & \hdots & \alpha \\
 \alpha & d_2 & \hdots & \alpha \\
 \vdots & \vdots & \ddots & \vdots \\
 \alpha & \alpha & \hdots & d_M \end{bmatrix} \right) = 0,
\end{equation}
with diagonal element $d_k = e\,c_{ik} R_k - m - (M-1) \alpha$.  Hence, the presence of species $i$ in the metacommunity at equilibrium imposes a constraint on the set of amounts of resources $R_k$.  Repeating this argument for other species $j$, we obtain additional constraints on the set $R_k$, $k=1,\ldots,M$.  Because there are $M$ amounts of resource $R_k$, it is generically possible to satisfy $M$ constraints.  As a result, there can be at most $M$ species present in the metacommunity at equilibrium.

\newpage

%%%%%%%%%%%%%%%%%%%%%%%%%%%%%%%%%%%%%%%%%%%%%%%%%%%%%%%%%%%%%%%%%%%%%%%%%%%%%%%%%%%%%%%
\section*{Appendix S2 \ Model reduction for large dispersal}
%%%%%%%%%%%%%%%%%%%%%%%%%%%%%%%%%%%%%%%%%%%%%%%%%%%%%%%%%%%%%%%%%%%%%%%%%%%%%%%%%%%%%%%

\subsection*{Theory} %%%%%%%%%%%%%%%%%%%%%%%%%%%%%%%%%%%%%%%%%%%%%%%%%%%%%%%%%%%%%%%%

We explain the procedure to study the large dispersal limit.  The procedure can be applied both to consumers and resources.  We consider a dynamical system for a mass quantity $X$ (like amount of resource or species biomass).  The system is spatially structured into $M$ patches.  The dynamical variables are the mass quantities $X_k$ in patch $k$ for $k=1,\ldots,M$.  The dynamical equations are
\begin{equation}
 \dyn{X_k} = f_k(X_k) + \gamma \big( \avg{X}-X_k \big).
 \label{eq:dynmass}
\end{equation}
with $\avg{X} = \frac{1}{M} \sum_k X_k$.  The first term represents the local dynamics (resource supply and consumption, or biomass growth and mortality); the second term represents the dispersal process with rate $\gamma$ (we have assumed uniform dispersal, but the model reduction also works for non-uniform dispersal).  Compared to dynamical system (1), both the dynamics of the resource and the dynamics of a consumer species are of the form (\ref{eq:dynmass}).

First we note that the dispersal process conserves total mass (that is, total amount of resource or total species biomass).  The dispersal process redistributes the masses $X_k$ over the different patches without changing the total mass $\sum_k X_k$.  Mathematically, this can be seen in the dynamical equation for $\avg{X}$,
\begin{equation}
 \dyn{\avg{X}} = \frac{1}{M} \sum_k f_k(X_k),
 \label{eq:avgmass}
\end{equation}
which does not contain dispersal terms.  In the limit of large dispersal, $\gamma\to\infty$, the dispersal terms dominate the local dynamics terms in (\ref{eq:dynmass}).  This implies that the relative mass in different patches is determined by the dispersal process, and is not affected by the local dynamics (the total mass in the system is not affected by the dispersal process, see equation (\ref{eq:avgmass})).  Mathematically, this corresponds to the quasi-stationary approximation:  we assume that the dispersal process reaches equilibrium before the local dynamics modify local mass $X_k$.  From equation (\ref{eq:dynmass}) it follows that the dispersal equilibrium is given by $X_k = \avg{X}$.  Substituting this equilibrium in (\ref{eq:avgmass}), we get
\begin{equation}
 \dyn{\avg{X}} = \frac{1}{M} \sum_k f_k(\avg{X}).
 \label{eq:redmass}
\end{equation}
The right-hand side of this dynamical equation does not depend on local mass $X_k$, but only on average mass $\avg{X}$.  Hence, using the quasi-stationary approximation we have replaced the set of $M$ equations for $X_k$ by a single equation for $\avg{X}$.  

The model reduction (\ref{eq:redmass}) also works if the dynamics (\ref{eq:dynmass}) are coupled to other dynamical equations.  It suffices to replace every occurrence of $X_k$ in the coupled dynamical equations with average mass $\avg{X}$.  Moreover, several model reductions can be applied successively.  For example, once we have applied the model reduction for one consumer species, we can repeat the same procedure and apply the model reduction for another consumer species or for the resource.

\subsection*{Application to model~(1)} %%%%%%%%%%%%%%%%%%%%%%%%%%%%%%%%%%%%%%%%%%%%%%

First, we construct the reduction of model~(1) for $\alpha\to\infty$.  We apply the procedure of the previous section with
\begin{equation*}
 \begin{aligned}
 X_k &\longrightarrow N_{ik} \\
 \avg{X} &\longrightarrow \avg{N_i} \\
 f_k(X_k) &\longrightarrow e\,c_{ik} R_k N_{ik} - m N_{ik}.
 \end{aligned}
\end{equation*}
The reduced dynamics (\ref{eq:redmass}) are
\begin{equation}
 \dyn{\avg{N_i}} = \frac{1}{M} \sum_k e\,c_{ik} R_k \avg{N_i} - m \avg{N_i}.
\end{equation}
Replacing $N_{ik}$ by $\avg{N_i}$ in the dynamics of $R_k$, we find
\begin{equation}
 \begin{aligned}
 \dyn{\avg{N_i}} &= \sum_k \frac{e\,c_{ik}}{M} R_k \avg{N_i} - m \avg{N_i} \\
 \dyn{R_k} &= g_k(R_k) - \sum_i c_{ik} R_k \avg{N_i} + \beta M \big( \avg{R} - R_k \big).
 \end{aligned}
 \label{eq:redalpha}
\end{equation}
We get reduced model~(2) by setting $\beta=0$.

Second, we construct the reduction of model~(1) for $\beta\to\infty$.  We apply the procedure of the previous section with
\begin{equation*}
 \begin{aligned}
 X_k &\longrightarrow R_k \\
 \avg{X} &\longrightarrow \avg{R} \\[-3pt]
 f_k(X_k) &\longrightarrow g_k(R_k) - \sum_i c_{ik} R_k N_{ik}.
 \end{aligned}
\end{equation*}
The reduced dynamics (\ref{eq:redmass}) are
\begin{equation}
 \dyn{\avg{R}} = \frac{1}{M} \sum_k g_k(\avg{R}) - \frac{1}{M} \sum_k c_{ik} \avg{R} N_{ik}.
\end{equation}
Replacing $R_k$ by $\avg{R}$ in the dynamics of $N_{ik}$, we find
\begin{equation}
 \begin{aligned}
 \dyn{N_{ik}} &= e\,c_{ik} \avg{R} N_{ik} - m N_{ik} + \alpha M \big( \avg{N_i} - N_{ik} \big) \\
 \dyn{\avg{R}} &= G(\avg{R}) - \sum_{i,k} \frac{c_{ik}}{M} \avg{R} N_{ik}.
 \end{aligned}
 \label{eq:redbeta}
\end{equation}
We get reduced model~(3) by setting $\alpha=0$.

Third, we construct the reduction of model~(1) for $\alpha\to\infty$ and $\beta\to\infty$.  We can take the limit $\beta\to\infty$ of model~(\ref{eq:redalpha}) or the limit $\alpha\to\infty$ of model~(\ref{eq:redbeta}).  Both methods lead to the same result.  Here we take the limit $\beta\to\infty$ of model~(\ref{eq:redalpha}).  Using
\begin{equation*}
 \begin{aligned}
 X_k &\longrightarrow R_k \\
 \avg{X} &\longrightarrow \avg{R} \\[-3pt]
 f_k(X_k) &\longrightarrow g_k(R_k) - \sum_i c_{ik} R_k \avg{N_i},
 \end{aligned}
\end{equation*}
we apply the procedure of the previous section to get
\begin{equation}
 \dyn{\avg{R}} = G(\avg{R}) - \frac{1}{M} \sum_{i,k} c_{ik} \avg{R} \avg{N_i}.
\end{equation}
Replacing $R_k$ by $\avg{R}$ in the dynamics of $\avg{N_i}$, we find
\begin{equation}
 \begin{aligned}
 \dyn{\avg{N_i}} &= \sum_k \frac{e\,c_{ik}}{M} \avg{R} \avg{N_i} - m \avg{N_i} \\
 \dyn{\avg{R}} &= G(\avg{R}) - \sum_{i,k} \frac{c_{ik}}{M} \avg{R} \avg{N_i},
 \end{aligned}
\end{equation}
which is reduced model~(4).

\newpage

%%%%%%%%%%%%%%%%%%%%%%%%%%%%%%%%%%%%%%%%%%%%%%%%%%%%%%%%%%%%%%%%%%%%%%%%%%%%%%%%%%%%%%%
\section*{Appendix S3 \ Quantifying metacommunity diversity}
%%%%%%%%%%%%%%%%%%%%%%%%%%%%%%%%%%%%%%%%%%%%%%%%%%%%%%%%%%%%%%%%%%%%%%%%%%%%%%%%%%%%%%%

In this study we focus on patterns of metacommunity diversity.  Diversity is most easily expressed in terms of number of species.  In a metacommunity context we distinguish local species richness and regional species richness.  Local species richness $S_\text{loc}$ can be defined as
\begin{equation}
 S_\text{loc} = \sum_k p^{(k)} S_\text{loc}^{(k)},
 \label{eq:defsloc}
\end{equation}
with $S_\text{loc}^{(k)}$ the number of species in local community $k$,
\[
 S_\text{loc}^{(k)} = \#\big\{ i \,\big|\, N_{ik} > 0 \big\},
\]
and $p^{(k)}$ the fraction of metacommunity biomass present in local community $k$,
\[
 p^{(k)} = \frac{\sum_i N_{ik}}{\sum_{i,k} N_{ik}}.
\]
We prefer using a weighted average in (\ref{eq:defsloc}) to moderate the contribution of local communities with small biomass.  Regional species richness $S_\text{reg}$ can be defined as the number of species in the metacommunity,
\begin{equation}
 S_\text{reg} = \#\big\{ i \,\big|\, \sum_k N_{ik} > 0 \big\}.
 \label{eq:defsreg}
\end{equation}
Definitions (\ref{eq:defsloc}) and (\ref{eq:defsreg}), however, have a major problem.  If consumer dispersal $\alpha$ is not zero, local and regional diversity are equal (see Figure~S1, row (a)).  Indeed, if a species is present in one of the local communities, it will be dispersed to all other local communities.  Hence, it will contribute to each $S_\text{loc}^{(k)}$ and to $S_\text{loc}$.  As a result, definitions (\ref{eq:defsloc}) and (\ref{eq:defsreg}) are not appropriate to distinguish diversity at the local and the regional scale.

The latter problem can be solved by introducing a biomass threshold $\theta$.  A species is considered to be present in a local community only if its biomass exceeds threshold $\theta$.  Local species richness $S_\text{loc}(\theta)$ is then defined as
\begin{equation}
 S_\text{loc}(\theta) = \sum_k p^{(k)} S_\text{loc}^{(k)}(\theta),
 \label{eq:defslocth}
\end{equation}
with $S_\text{loc}^{(k)}(\theta)$ the number of species in local community $k$ with biomass larger than $\theta$,
\[
 S_k(\theta) = \#\big\{ i \,\big|\, N_{ik} > \theta \big\}.
\]
Regional species richness $S_\text{reg}(\theta)$ is defined as the number of species in the metacommunity with biomass larger than $\theta$,
\begin{equation}
 S_\text{reg}(\theta) = \#\big\{ i \,\big|\, \sum_k N_{ik} > \theta \big\}.
 \label{eq:defsregth}
\end{equation}
Using definitions (\ref{eq:defslocth}) and (\ref{eq:defsregth}) local and regional diversity are equal only for large consumer dispersal $\alpha$ (see Figure~S1, row (b)).  Patterns for local and regional diversity are qualitatively different.

Another solution to the problem consists in using Shannon diversity instead of species richness to quantify metacommunity diversity.  Local diversity $D_\text{loc}$ is then defined as
\begin{equation}
 D_\text{loc} = \sum_k p^{(k)} D_\text{loc}^{(k)},
 \label{eq:defdloc}
\end{equation}
where $D_\text{loc}^{(k)}$ is the Shannon diversity of local community $k$,
\[
 D_\text{loc}^{(k)} = -\sum_i p_i^{(k)} \ln p_i^{(k)},
\]
and $p_i^{(k)}$ the relative biomass of species $i$ in local community $k$,
\[
 p_i^{(k)} = \frac{N_{ik}}{\sum_i N_{ik}}.
\]
The weighted average in (\ref{eq:defslocth}) allows us to moderate the contribution of local communities with small biomass.  Regional diversity $D_\text{reg}$ is defined as the regional Shannon diversity index,
\begin{equation}
 D_\text{reg} = - \sum_i p_i \ln p_i,
 \label{eq:defdreg}
\end{equation}
with $p_i$ the relative biomass of species $i$ in the metacommunity,
\[
 p_i = \frac{\sum_k N_{ik}}{\sum_{i,k} N_{ik}}.
\]
Definitions (\ref{eq:defdloc}) and (\ref{eq:defdreg}) lead to metacommunity diversity patterns that are similar to those obtained with definitions (\ref{eq:defslocth}) and (\ref{eq:defsregth}) (see Figure~S1, row (c)).  We prefer using Shannon diversity because it facilitates the comparison of results for two-patch and five-patch metacommunities.  Furthermore, it does not require an additional parameter $\theta$.

\newpage

%%%%%%%%%%%%%%%%%%%%%%%%%%%%%%%%%%%%%%%%%%%%%%%%%%%%%%%%%%%%%%%%%%%%%%%%%%%%%%%%%%%%%%%
\section*{Appendix S4 \ Analysis of four limiting cases}
%%%%%%%%%%%%%%%%%%%%%%%%%%%%%%%%%%%%%%%%%%%%%%%%%%%%%%%%%%%%%%%%%%%%%%%%%%%%%%%%%%%%%%%

We present a detailed analysis of the four limiting cases of model~(1) as represented in Figure~1.  The analysis for large consumer dispersal $\alpha$ and/or large resource dispersal $\beta$ is based on the model reduction of Appendix~S2.  We are especially interested in local diversity~$D_\text{loc}$ and regional diversity~$D_\text{reg}$ defined in Appendix~S3.  

\subsection*{Case $\alpha=0$ and $\beta=0$} %%%%%%%%%%%%%%%%%%%%%%%%%%%%%%%%%%%%%%%

Dropping the dispersal terms in model~(1), the dynamics inside a patch are decoupled from the other patches.  At equilibrium, if species $i$ is present in patch $k$, the amount of resource in patch $k$ is equal to
\[
 R_k = \frac{m}{e\,c_{ik}}.
\]
Hence, a necessary condition for species $i$ to be present in patch $k$ is
\begin{equation}
 c_{ik} >
 \begin{cases}
  \frac{m}{e\,A_k} & \text{for an abiotic resource} \\
  \frac{m}{e\,B_k} & \text{for a biotic resource.}
 \end{cases}
 \label{eq:case1surv}
\end{equation}
At equilibrium the species with the largest $c_{ik}$ excludes the other species from local community $k$.  We denote this species by $j$.  If condition (\ref{eq:case1surv}) is satisfied for species $j$, the equilibrium in patch $k$ is
\begin{equation}
 \begin{aligned}
  R_k &= \frac{m}{e\,c_{jk}} \\
  N_{jk} &= \frac{e}{m}\, g_k(R_k) \\
  N_{ik} &= 0 \qquad \text{for all $i \neq j$.}
 \end{aligned}
\end{equation}
If condition (\ref{eq:case1surv}) is not satisfied for species $j$, the equilibrium in patch $k$ is
\begin{equation}
 \begin{aligned}
  R_k &=
  \begin{cases}
   A_k & \text{for an abiotic resource} \\
   B_k & \text{for a biotic resource}
  \end{cases} \\
  N_{ik} &= 0 \qquad \text{for all $i$.}
 \end{aligned}
\end{equation}

Because each patch is occupied by (at most) one species, local diversity $D_\text{loc}$ is zero.  Because different species can occupy different patches, regional diversity $D_\text{reg}$ can be large.

\subsection*{Case $\alpha\to\infty$ and $\beta=0$} %%%%%%%%%%%%%%%%%%%%%%%%%%%%%%%

The reduced model~(2) describes $S$ consumer species competing for $M$ effective resources.  An effective resource corresponds to the resource isolated in a patch.  Analytical formulas to determine which species coexist at equilibrium are complicated.  However, non-spatial consumer-resource theory can be applied to determine graphically the set of coexisting species.  This is illustrated in Figure~S2 for the metacommunities of Figure~2.

Explicit formulas can be obtained for some special cases.  First, consider an ``extreme specialist'' species $i$.  That is, species $i$ consumes the resource in patch $k$ at rate $c_{ik}$, but it cannot consume the resource in the other patches.  At equilibrium, if species $i$ is present in the metacommunity, the amount of resource in patch $k$ is equal to
\[
 R_k = \frac{M\,m}{e\,c_{ik}}.
\]
Hence, a necessary condition for an extreme specialist species $i$ to be present in the metacommunity is
\begin{equation}
 c_{ik} >
 \begin{cases}
  \frac{M\,m}{e\,A_k} & \text{for an abiotic resource} \\
  \frac{M\,m}{e\,B_k} & \text{for a biotic resource.}
 \end{cases}
\end{equation}

Second, consider an ``extreme generalist'' species $i$.  That is, species $i$ consumes the resource in all patches at the same rate $c_{ik} = c_i$.  At equilibrium, if species $i$ is present in the metacommunity, the average amount of resource is equal to
\[
 \avg{R} = \frac{m}{e\,c_i}.
\]
Hence, a necessary condition for an extreme generalist species $i$ to be present in the metacommunity is
\begin{equation}
 c_i >
 \begin{cases}
  \frac{m}{e\,\avg{A}} & \text{for an abiotic resource} \\
  \frac{m}{e\,\avg{B}} & \text{for a biotic resource.}
 \end{cases}
\end{equation}

Because each patch has the same community composition, we have $D_\text{loc} = D_\text{reg}$, which can be small or large depending on the competition outcome.

\subsection*{Case $\alpha=0$ and $\beta\to\infty$} %%%%%%%%%%%%%%%%%%%%%%%%%%%%%%%

The reduced model~(3) describes $M S$ effective consumers competing for a single resource.  An effective consumer corresponds to a consumer species isolated in a patch.  At equilibrium, if species $i$ is present in patch $k$, the amount of resource is equal to
\[
 R_k = \avg{R} = \frac{m}{e\,c_{ik}}.
\]
Hence, a necessary condition for species $i$ to be present in patch $k$ is
\begin{equation}
 c_{ik} >
 \begin{cases}
  \frac{m}{e\,\avg{A}} & \text{for an abiotic resource} \\
  \frac{m}{e\,\avg{B}} & \text{for a biotic resource.}
 \end{cases}
 \label{eq:case3surv}
\end{equation}
At equilibrium the species-patch combination with the largest $c_{ik}$ excludes the other species from the metacommunity.  We denote this species-patch combination by $(j,\ell)$.  If condition (\ref{eq:case3surv}) is satisfied for species-patch combination $(j,\ell)$, the equilibrium is
\begin{equation}
 \begin{aligned}
  R_k &= \avg{R} = \frac{m}{e\,c_{j\ell}} \qquad \text{for all $k$} \\
  N_{j\ell} &= M \frac{e}{m}\,G(\avg{R}) \\
  N_{ik} &= 0 \qquad \text{for all $(i,k) \neq (j,\ell)$.}
 \end{aligned}
\end{equation}
If condition (\ref{eq:case3surv}) is not satisfied for species-patch combination $(j,\ell)$, the equilibrium is
\begin{equation}
 \begin{aligned}
  R_k &= \avg{R} =
  \begin{cases}
   \avg{A} & \text{for an abiotic resource} \\
   \avg{B} & \text{for a biotic resource}
  \end{cases} \qquad \text{for all $k$} \\
  N_{ik} &= 0 \qquad \text{for all $(i,k)$.}
 \end{aligned}
\end{equation}

Because (at most) one species is present in the metacommunity, we have $D_\text{loc} = D_\text{reg} = 0$.

\subsection*{Case $\alpha\to\infty$ and $\beta\to\infty$} %%%%%%%%%%%%%%%%%%%%%%%

The reduced model~(4) describes $S$ consumer species competing for a single resource.  At equilibrium, if species $i$ is present in the metacommunity, the amount of resource is equal to
\[
 R_k = \avg{R} = \frac{m}{e\,\avg{c_i}}.
\]
Hence, a necessary condition for species $i$ to be present in the metacommunity is
\begin{equation}
 \avg{c_i} >
 \begin{cases}
  \frac{m}{e\,\avg{A}} & \text{for an abiotic resource} \\
  \frac{m}{e\,\avg{B}} & \text{for a biotic resource.}
 \end{cases}
 \label{eq:case4surv}
\end{equation}
At equilibrium the species with the largest $\avg{c_i}$ excludes the other species from the metacommunity.  We denote this species by $j$.  If condition (\ref{eq:case4surv}) is satisfied for species $j$, the equilibrium is
\begin{equation}
 \begin{aligned}
  R_k &= \avg{R} = \frac{m}{e\,\avg{c_j}}
  \qquad \text{for all $k$} \\
  N_{jk} &= \avg{N_j} = \frac{e}{m}\,G(\avg{R})
  \qquad \text{for all $k$} \\
  N_{ik} &= \avg{N_i} = 0
  \qquad \text{for all $i \neq j$ and all $k$.}
 \end{aligned}
\end{equation}
If condition (\ref{eq:case4surv}) is not satisfied for species $j$, the equilibrium is
\begin{equation}
 \begin{aligned}
  R_k &= \avg{R} =
  \begin{cases}
   \avg{A} & \text{for an abiotic resource} \\
   \avg{B} & \text{for a biotic resource}
  \end{cases} \qquad \text{for all $k$} \\
  N_{ik} &= \avg{N_i} = 0
  \qquad \text{for all $(i,k)$}
 \end{aligned}
\end{equation}

Because (at most) one species is present in the metacommunity, we have $D_\text{loc} = D_\text{reg} = 0$.

\newpage

%%%%%%%%%%%%%%%%%%%%%%%%%%%%%%%%%%%%%%%%%%%%%%%%%%%%%%%%%%%%%%%%%%%%%%%%%%%%%%%%%%%%%%%
\section*{Appendix S5 \ Generating simulation parameters}
%%%%%%%%%%%%%%%%%%%%%%%%%%%%%%%%%%%%%%%%%%%%%%%%%%%%%%%%%%%%%%%%%%%%%%%%%%%%%%%%%%%%%%%

We construct a procedure to study the generic patterns of metacommunity diversity as predicted by model~(1).  This procedure, based on randomly generating parameter values of model~(1), is used to obtain the simulation results for the five-patch metacommunities shown in Figure~4 and Figures~S8--S11.

We have to specify the parameters $c_{ik}$ describing the spatial preference of each species $i$ and the patch fertilities $A_k$ or $B_k$ (abiotic or biotic resource) describing the spatial dependence of the resource.  Other parameter can be chosen as $e = m = 1$ and $a = 1$ or $b = 1$ (abiotic or biotic resource) without loss of generality.

A straightforward procedure uses randomly chosen, mutually independent parameters.  For example, we can draw consumption rates $c_{ik}$ independently from an exponential distribution (see Figure~S7, panel (a)) and patch fertilities $A_k$ or $B_k$ independently from an exponential distribution.  An example of diversity patterns obtained with this procedure is shown in Figure~S8, row (a).  The patterns are very similar to those obtained for the two-patch metacommunities (compare with Figure~3, row (d), for example).  The corresponding one-dimensional diversity-dispersal relationships are shown in Figure~S9, row (a).

The procedure with independent consumption rates can generate species that are among the best consumers in each of the patches.  There is no trade-off between consumption rates $c_{ik}$ of species $i$ in different patches $k$.  In particular, the same species can be the best specialist consumer in one or more patches and the best generalist.  Conversely, species can have small consumption rates in each of the patches, so that they are absent in the metacommunity at equilibrium for any dispersal $\alpha$ and $\beta$ and patch fertility $A_k$ or $B_k$.

We modify the procedure with independently drawn parameters by imposing a trade-off between consumption rates $c_{ik}$ of species $i$ in different patches $k$.  We require that
\begin{equation}
 \sum_k c_{ik} = C \qquad \text{for all species $i$.}
 \label{eq:lintraoff}
\end{equation}
We implement this constraint by drawing the consumption rates $c_{ik}$ independently from an exponential distribution.  If constraint~(\ref{eq:lintraoff}) is satisfied (allowing a small deviation), we accept the chosen parameters.  If constraint~(\ref{eq:lintraoff}) is not satisfied, we draw another set of consumption rates $c_{ik}$, until the constraint is satisfied (see Figure~S7, panel (b)).  The patch fertilities $A_k$ or $B_k$ are drawn independently from an exponential distribution.

An example of diversity patterns obtained with this procedure is shown in Figure~S8, row (b) and Figure~S9, row (b).  The patterns are similar to those obtained previously for the two-patch and five-patch metacommunities.  However, there is a difference for large $\alpha$ and large $\beta$.  Several species coexist locally for relatively large values of $\alpha$ and $\beta$, where previously the best generalist species dominated the metacommunity.  This difference is due to the linearity of constraint~(\ref{eq:lintraoff}), which implies a very small competitive difference between species for large $\alpha$ and large $\beta$.  In fact, constraint (\ref{eq:lintraoff}) does not impose a trade-off between specialists and generalists.  Even if a species is the best specialist in one of the patches, it is as good a generalist as all other species.

To impose a trade-off between specialists and generalists we have to replace the linear constraint~(\ref{eq:lintraoff}) with a nonlinear one.  A convex trade-off like
\begin{equation}
 \sum_k \sqrt{c_{ik}} = C \qquad \text{for all species $i$,}
\end{equation}
would increase the advantage of specialists with respect to generalists.  The metacommunity, even for large $\alpha$ and large $\beta$, would be dominated by specialist consumers.  We need a concave trade-off like
\begin{equation}
 \sum_k c_{ik}^2 = C \qquad \text{for all species $i$,}
 \label{eq:quadtraoff}
\end{equation}
to allow generalist species to coexist with and replace specialist species when increasing $\alpha$ and $\beta$.  We use the same procedure as above except that we replace constraint~(\ref{eq:lintraoff}) with constraint~(\ref{eq:quadtraoff}) (see Figure~S7, panel (c)).

An example of diversity patterns obtained with this procedure is shown in Figure~S8, row (c) and Figure~S9, row (c).  The patterns are very similar to those obtained previously for the two-patch and five-patch metacommunities.  For small $\alpha$ the specialist species in the species pool occupy the metacommunity; for large $\alpha$ the generalist species tend to occupy the metacommunity.  Four more metacommunities with trade-off~(\ref{eq:quadtraoff}) but with a biotic resource are shown in Figures~S10 and S11.  The diversity patterns differ only for large $\alpha$ and small $\beta$, where diversity can be small (metacommunity (a)) or large (metacommunity (b))).  Again, the patterns are very similar to those obtained previously.

Finally, we give the parameter values used in Figure~4.  Consumption rates are drawn randomly with a quadratic trade-off given by $\sum_k c_{ik}^2 = 100 \pm 1$.  The resource is biotic with $b = 1$ and patch fertilities $B_k$ drawn independently from an exponential distribution with mean 2.  Other parameters are $e = 1$ and $m = 1$.

\newpage

%%%%%%%%%%%%%%%%%%%%%%%%%%%%%%%%%%%%%%%%%%%%%%%%%%%%%%%%%%%%%%%%%%%%%%%%%%%%%%%%%%%%%%%
\section*{Appendix S6 \ Comparison with Loreau et al.~(2003a)}
%%%%%%%%%%%%%%%%%%%%%%%%%%%%%%%%%%%%%%%%%%%%%%%%%%%%%%%%%%%%%%%%%%%%%%%%%%%%%%%%%%%%%%%

We discuss the connection between model (1) and the model studied in Loreau et al.~(2003a).  The latter model is similar to model~(1) except that the consumption rates $c_{ik}$ are time-dependent, an extinction threshold is implemented in the model dynamics, and resource dispersal $\beta$ is set to zero.  To establish a mathematical link between the two models, we note that in the model of Loreau et al.~(2003a) the extrinsic fluctuations are slow compared to the intrinsic dynamics (see below).  Hence, we can apply the quasi-stationary approximation:  we assume that the system reaches equilibrium before the environment changes.  We study the diversity-dispersal relationship predicted by Loreau et al.~(2003a) by fixing the consumption rates $c_{ik}$ at time $t=0$.

Loreau et al.~(2003a) consider a metacommunity with $M=7$ patches and a species pool with $S=7$ species.  The consumption rate $c_{ik}$ at time $t=0$ are 
\begin{equation}
 [ c_{ik} ] = \begin{pmatrix}
 1.50 & 1.33 & 1.17 & 1.00 & 0.83 & 0.67 & 0.50 \\
 1.33 & 1.50 & 1.33 & 1.17 & 1.00 & 0.83 & 0.67 \\
 1.17 & 1.33 & 1.50 & 1.33 & 1.17 & 1.00 & 0.83 \\
 1.00 & 1.17 & 1.33 & 1.50 & 1.33 & 1.17 & 1.00 \\
 0.83 & 1.00 & 1.17 & 1.33 & 1.50 & 1.33 & 1.17 \\
 0.67 & 0.83 & 1.00 & 1.17 & 1.33 & 1.50 & 1.33 \\
 0.50 & 0.67 & 0.83 & 1.00 & 1.17 & 1.33 & 1.50
 \end{pmatrix}.
 \label{eq:consrate1}
\end{equation}
Rows correspond to species and columns correspond to patches.  Other parameter values (using our notation) are $M=7$, $S=7$, $e = 0.2$, $m = 0.2$, $a = 10$, $A_k = 15$.  Figure~S12, panel (a) shows that the relationship between local diversity and dispersal $\alpha$ is hump-shaped, as reported in Loreau et al.~(2003a).  There are quantitative differences with the diversity-dispersal relationship shown in Loreau et al.~(2003a).  However, the approximation that the extrinsic fluctuations are slow compared to the intrinsic dynamics is justified for most dispersal values $\alpha$.  The time-scale of the extrinsic fluctuations (their period) is $40000$, whereas the time-scale of the intrinsic dynamics (the reciprocal of the real part of the dominant eigenvalue of the Jacobian) is $50$ (for $\alpha=0.0001$), 60 (for $\alpha=0.001$), 1500 (for $\alpha=0.01$), 400 (for $\alpha=0.1$) and 300 (for $\alpha=1$).

The hump-shaped diversity-dispersal relationship is consistent with the results of the main text.  Each species is the best competitor in one of the patches.  For small $\alpha$ and small $\beta$, species $i$ dominates patch $k=i$.  Each species has the same consumption rate in its preferred patch ($c_{ii} = 1.5$), leading to (non-generic) neutral coexistence for small $\alpha$ and large $\beta$.  The fourth species has a competitive advantage for large $\alpha$, because its average consumption rate $\avg{c_i}$ is larger than the other species.  Hence, it dominates the metacommunity for large $\alpha$ and large $\beta$.  For consumption rates (\ref{eq:consrate1}) the fourth species dominates the metacommunity also for large $\alpha$ and small $\beta$.

It is interesting to consider a set of slightly modified parameter values.  We keep the same parameters as above, but change the consumption rates (\ref{eq:consrate1}) to
\begin{equation}
 [ c_{ik} ] = \begin{pmatrix}
 1.50 & 0.50 & 0.45 & 0.40 & 0.35 & 0.30 & 0.25 \\
 0.50 & 1.50 & 0.50 & 0.45 & 0.40 & 0.35 & 0.30 \\
 0.45 & 0.50 & 1.50 & 0.50 & 0.45 & 0.40 & 0.35 \\
 0.40 & 0.45 & 0.50 & 1.50 & 0.50 & 0.45 & 0.40 \\
 0.35 & 0.40 & 0.45 & 0.50 & 1.50 & 0.50 & 0.45 \\
 0.30 & 0.35 & 0.40 & 0.45 & 0.50 & 1.50 & 0.50 \\
 0.25 & 0.30 & 0.35 & 0.40 & 0.45 & 0.50 & 1.50
 \end{pmatrix}.
 \label{eq:consrate2}
\end{equation}
These consumption rates have the same qualitative features as (\ref{eq:consrate1}).  Each species is specialized on a different patch and the fourth species is the best generalist.  The only difference with (\ref{eq:consrate1}) is that species are more specialized on their preferred patch.  Consequently, niche overlap between species is smaller, thereby facilitating species coexistence.  The diversity-dispersal relationship for (\ref{eq:consrate2}) is qualitatively different from the relationship for (\ref{eq:consrate1}):  it is increasing for most dispersal values $\alpha$ and reaches high diversity values for large $\alpha$, see Figure~S12, panel (b).  The fourth species does not dominate the metacommunity for large $\alpha$ and small $\beta$, but it coexists with the other species.  We conclude that the model of Loreau et al.~(2003a) predicts a richer set of diversity-dispersal relationships than only the hump-shaped relationship.

\clearpage\newpage

%%% FIGURE S1 %%%
\begin{figure}
\begin{center}
\includegraphics[width=.76\textwidth]{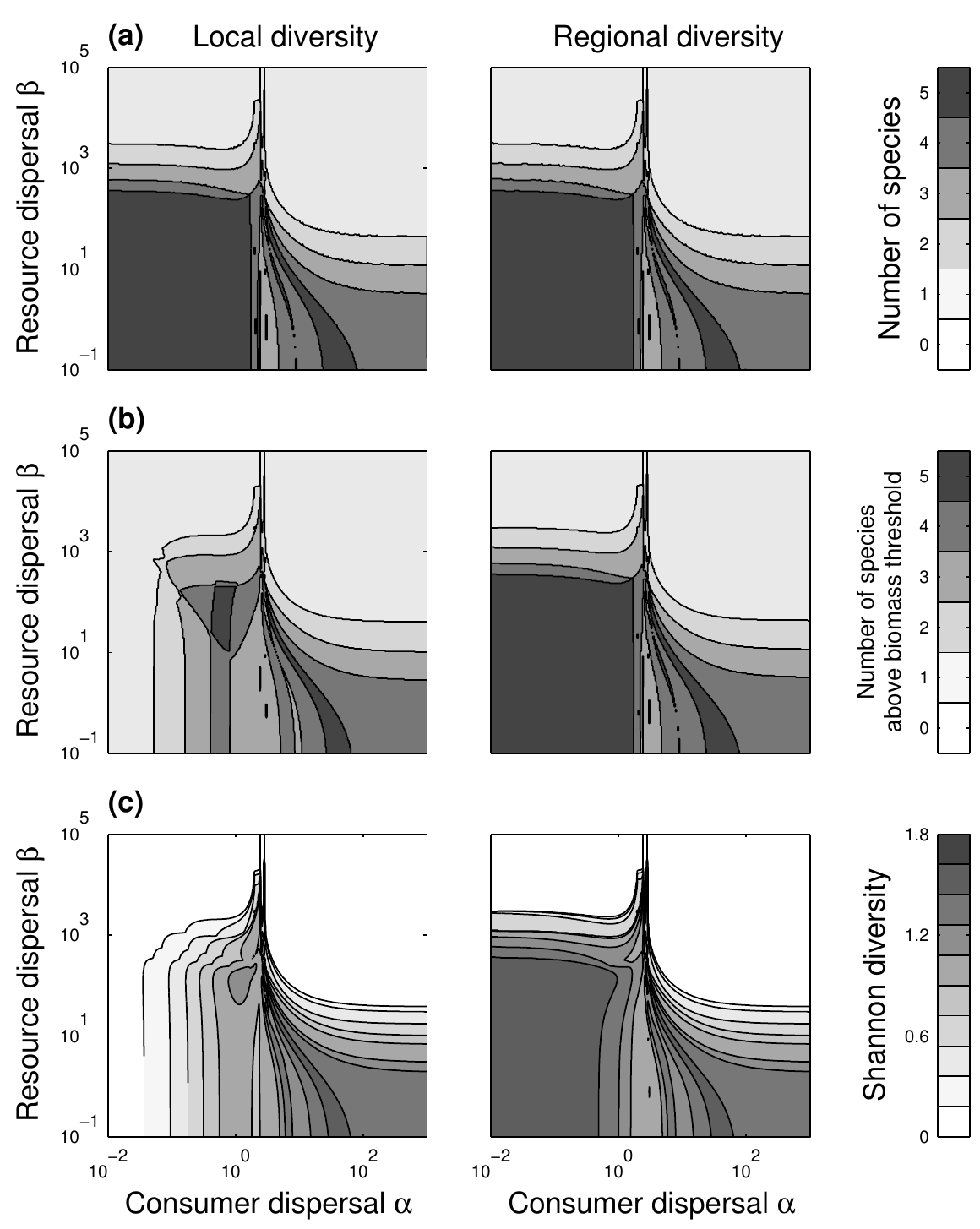}
\caption{Quantifying metacommunity diversity using species richness and Shannon diversity.  Same metacommunity as in Figure~4.  Panels~(a):  average number of species in local communities (left) and number of species in metacommunity (right).  Panels~(b):  average number of species with biomass larger than $\theta = 0.05$ in local communities (left) and number of species with biomass larger than $\theta = 0.05$ in metacommunity (right).  Panels~(c):  average Shannon diversity of local communities (left) and Shannon diversity of metacommunity (right).}
\end{center}
\end{figure}

\clearpage\newpage

%%% FIGURE S2 %%%
\begin{figure}
\begin{center}
\centerline{\includegraphics[width=.78\textwidth]{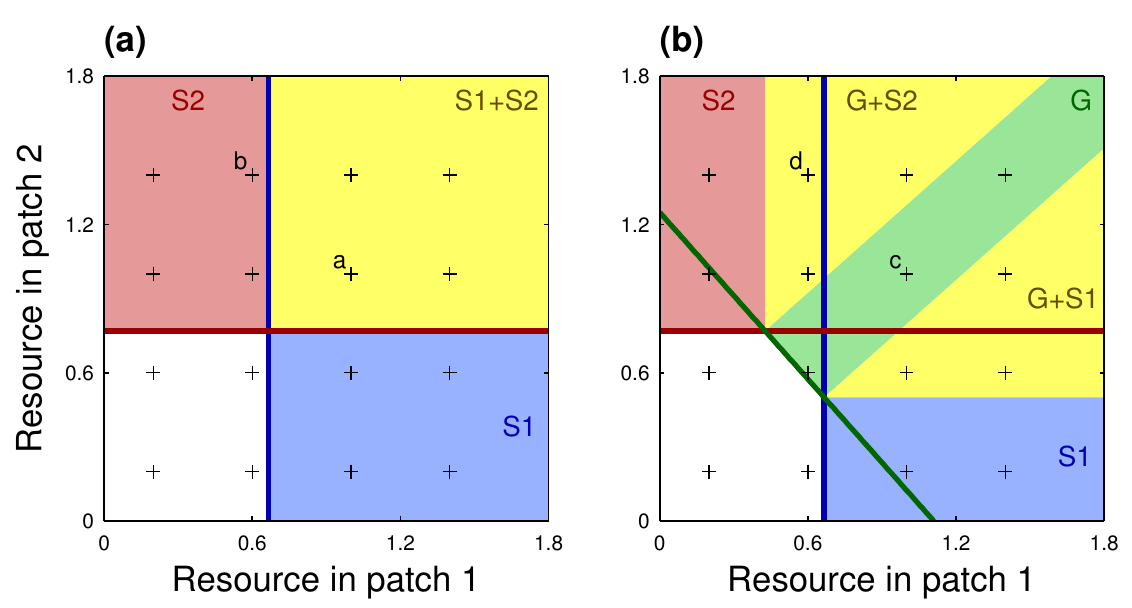}}
\caption{Effects of patch fertility on two-patch metacommunities for limiting case $\alpha\to\infty$ and $\beta=0$.  Non-spatial resource competition theory allows us to determine the metacommunity composition when $\alpha\to\infty$ and $\beta=0$.  Same metacommunities as in Figure~2.  In panel~(a) the species pool consists of two specialist species S1 and S2.  In panel~(b) the species pool consists of species S1 and S2 and generalist species G.  Dark blue, red and green lines are the zero net growth isoclines (ZNGIs).  Light blue, red, green and yellow regions are sets of vectors $(B_1,B_2)$ for which the indicated species persist;  no species persists in the white regions.  Black crosses indicate vectors $(B_1,B_2)$ used in other figures:  crosses in panel~(a) refer to panels of Figure~S3;  crosses in panel~(b) refer to panels of Figure~S4;  crosses with labels a, b, c and d refer to the four panels of Figure~2.}
\end{center}
\end{figure}

\clearpage\newpage

%%% FIGURE S3 %%%
\begin{figure}
\begin{center}
\centerline{\includegraphics[width=1.1\textwidth]{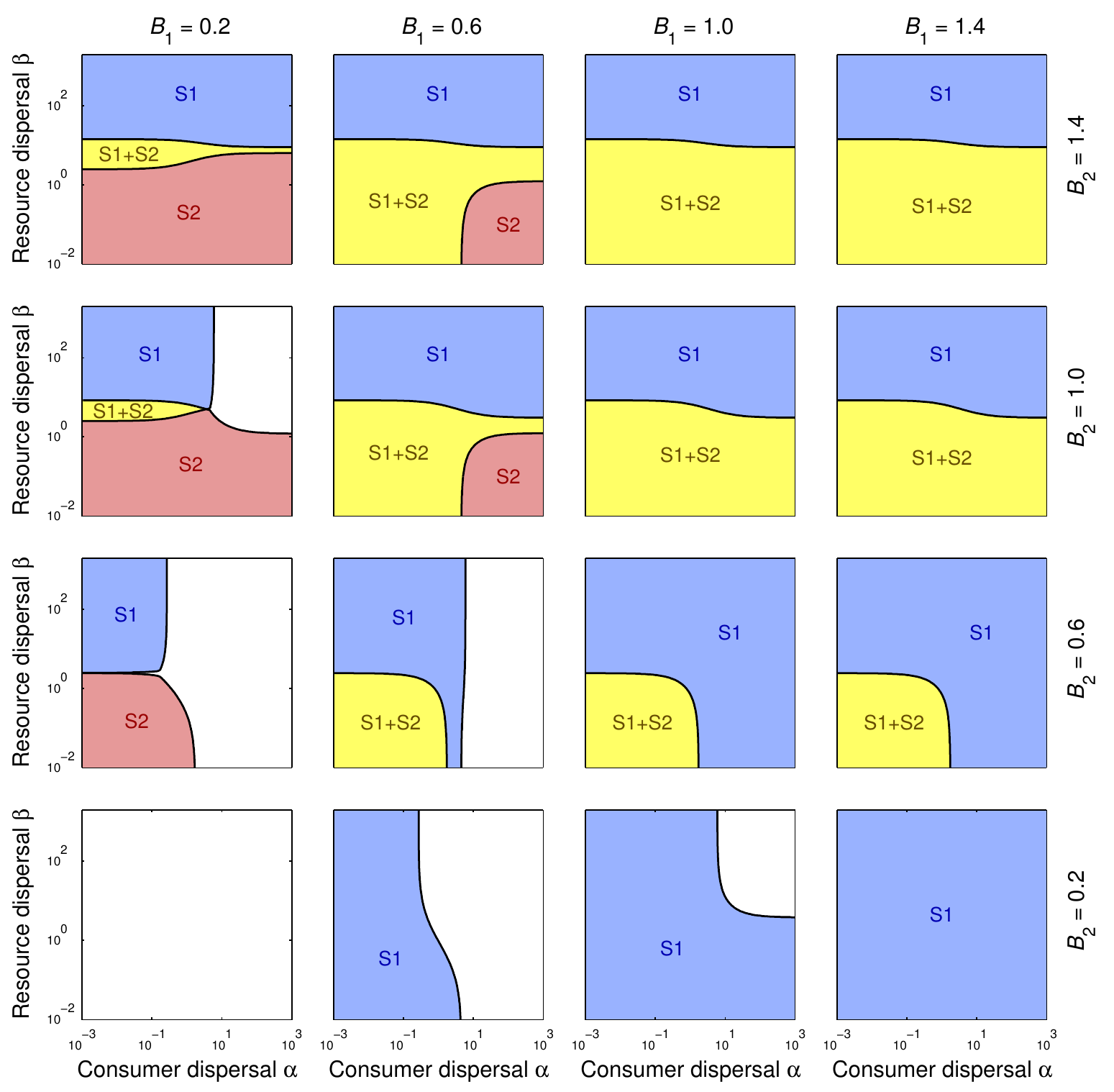}}
\caption{Effects of disperal and patch fertility on two-patch two-species metacommunity.  Same metacommunity as in Figure~2, panels (a--b).  The species pool consists of two specialist species.   Species S1 is specialized on patch 1 and species S2 is specialized on patch 2.  Dispersal rates $\alpha$ and $\beta$ vary within panels;  patch fertilities $B_1$ and $B_2$ vary within panels.  Light blue, red and yellow regions are sets of vectors $(\alpha,\beta)$ for which the indicated species persist;  no species persists in the white regions.}
\end{center}
\end{figure}

\clearpage\newpage

%%% FIGURE S4 %%%
\begin{figure}
\begin{center}
\centerline{\includegraphics[width=1.1\textwidth]{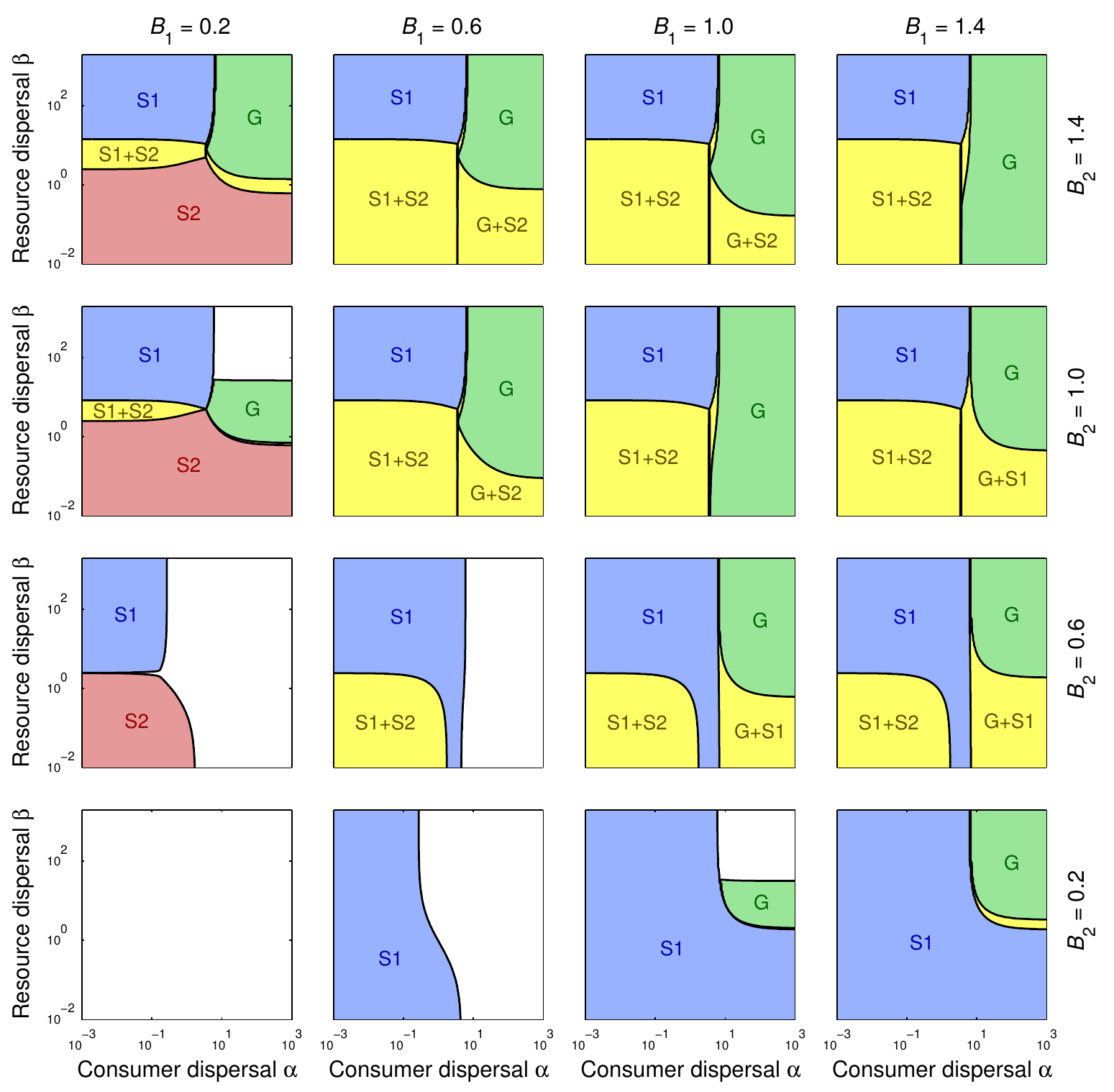}}
\caption{Effects of dispersal and patch fertility on two-patch three-species metacommunity.  Same metacommunity as in Figure~2, panels (c--d).  The species pool consists of specialist specices S1 and S2 and generalist species G.  Dispersal rates $\alpha$ and $\beta$ vary within panels;  patch fertilities $B_1$ and $B_2$ vary between panels.  Light blue, red, green and yellow regions are sets of vectors $(\alpha,\beta)$ for which the indicated species persist;  no species persists in the white regions.}
\end{center}
\end{figure}

\clearpage\newpage

%%% FIGURE S5 %%%
\begin{figure}
\begin{center}
\includegraphics[width=.96\textwidth]{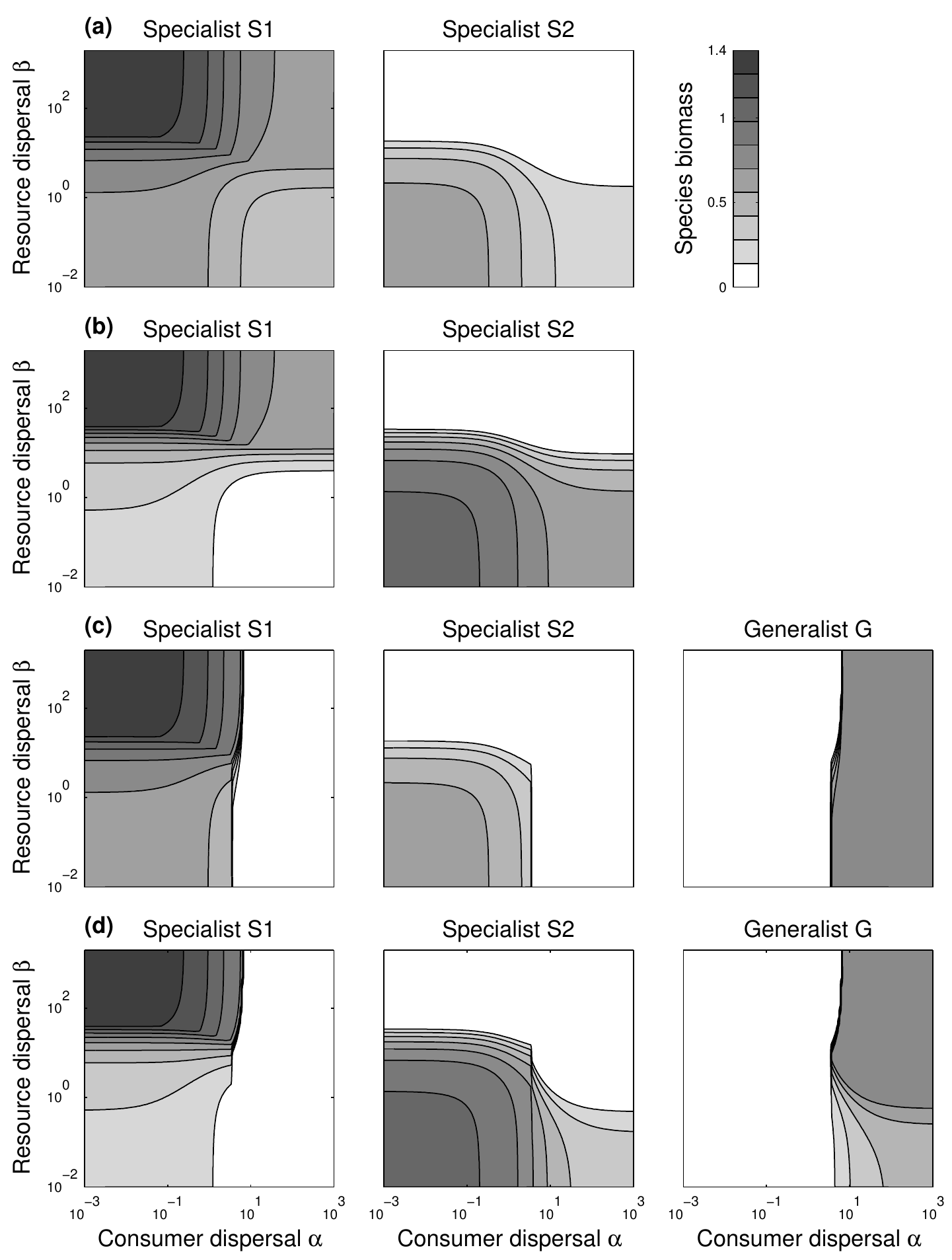}
\caption{Effects of consumer and resource dispersal on composition of two-patch metacommunity with abiotic resource.  Same as Figure~2 except that here the resouce is abiotic.  Parameter values for abiotic resource:  $a=1$; (a) $A_1=1$, $A_2=1$; (b) $A_1=0.6$, $A_2=1.4$; (c) $A_1=1$, $A_2=1$; (d) $A_1=0.6$, $A_2=1.4$.}
\end{center}
\end{figure}

\clearpage\newpage

%%% FIGURE S6 %%%
\begin{figure}
\begin{center}
\includegraphics[width=.76\textwidth]{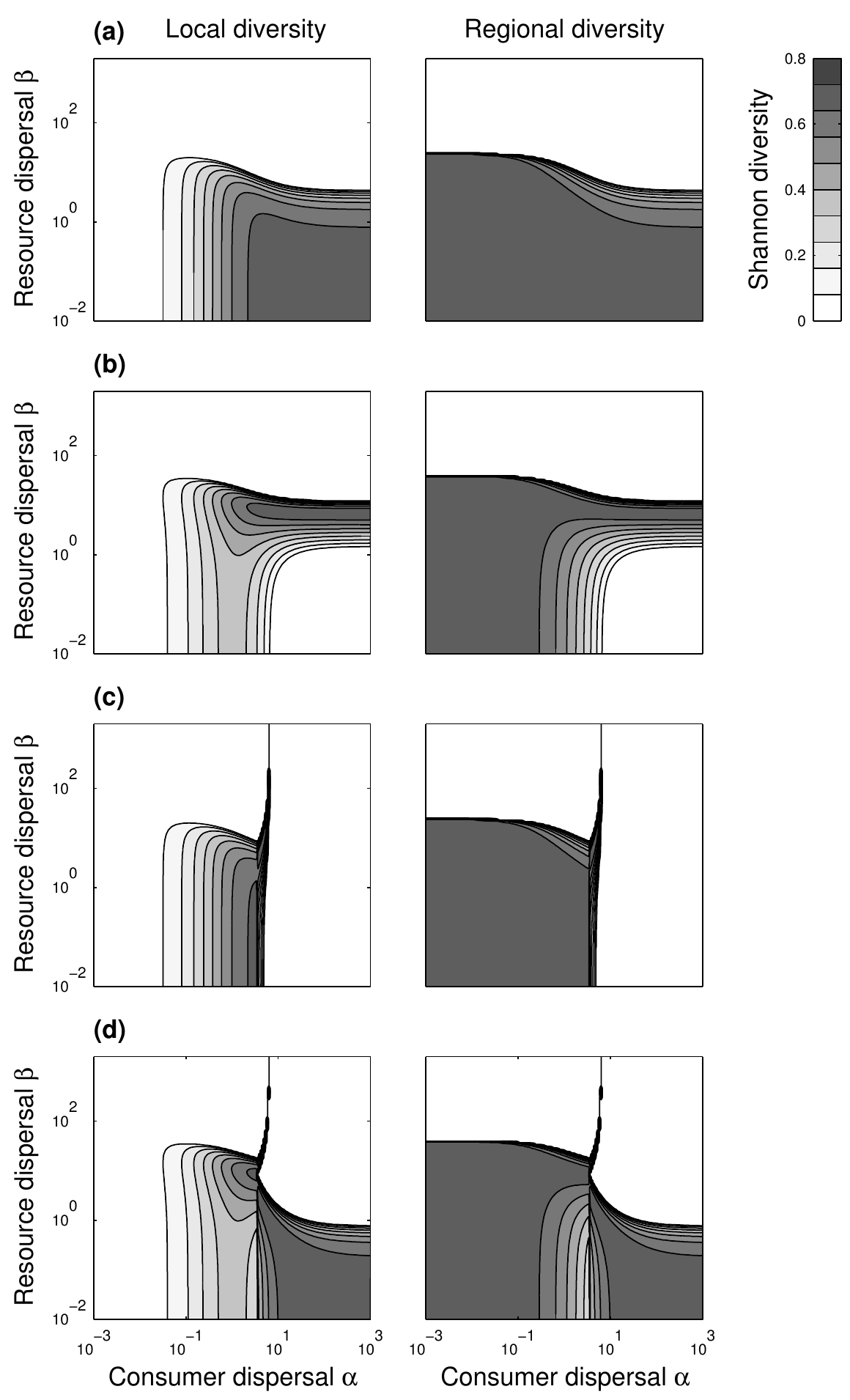}
\caption{Effects of consumer and resource dispersal on diversity of two-patch metacommunity with abiotic resource.  Same as Figure~3 except that here the resource is abiotic.  Parameter values for abiotic resource:  $a=1$; (a) $A_1=1$, $A_2=1$; (b) $A_1=0.6$, $A_2=1.4$; (c) $A_1=1$, $A_2=1$; (d) $A_1=0.6$, $A_2=1.4$.}
\end{center}
\end{figure}

\clearpage\newpage

%%% FIGURE S7 %%%
\begin{figure}
\begin{center}
\includegraphics[width=.9\textwidth]{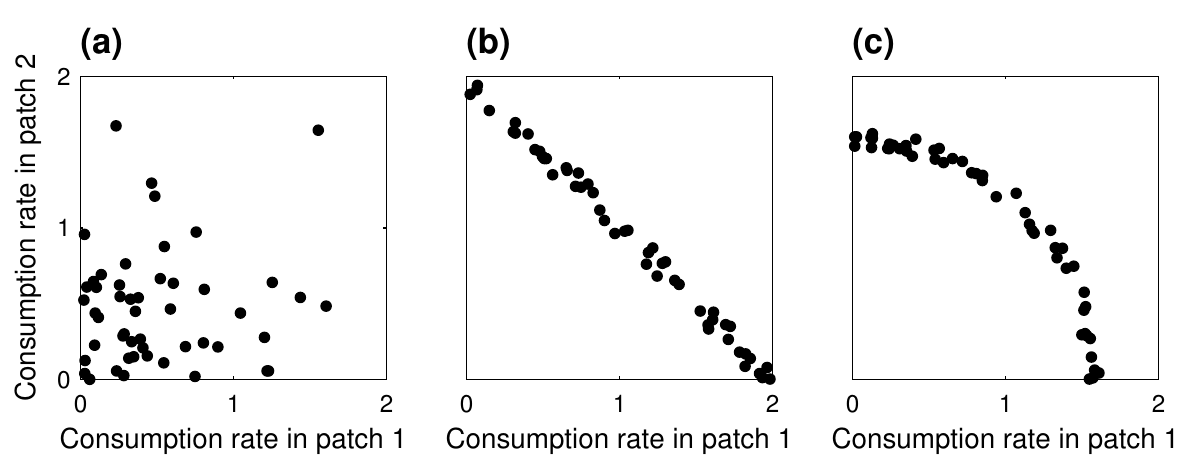}
\caption{Randomly generating species consumption rates.  Three different procedures are illustrated for a metacommunity with two patches and with a species pool consisting of 50 species.  A black dot represents the consumption rates $c_{i1}$ in patch 1 ($x$-axis) and $c_{i2}$ in patch 2 ($y$-axis) of a species $i$.  Panel (a):  Consumption rates $c_{ik}$ of species $i$ are drawn independently.  Here we use an exponential distribution with mean $0.5$.  Panel (b):  Consumption rates $c_{ik}$ for species $i$ are drawn randomly with a linear trade-off.  The trade-off curve is given by the equation $c_{i1} + c_{i2} = 2.0 \pm 0.1$.  Panel (c):  Consumption rates $c_{ik}$ for species $i$ are drawn randomly with a quadratic trade-off.  The trade-off curve is given by the equation $c_{i1}^2 + c_{i2}^2 = 2.5 \pm 0.2$.}
\end{center}
\end{figure}

\clearpage\newpage

%%% FIGURE S8 %%%
\begin{figure}
\begin{center}
\includegraphics[width=.76\textwidth]{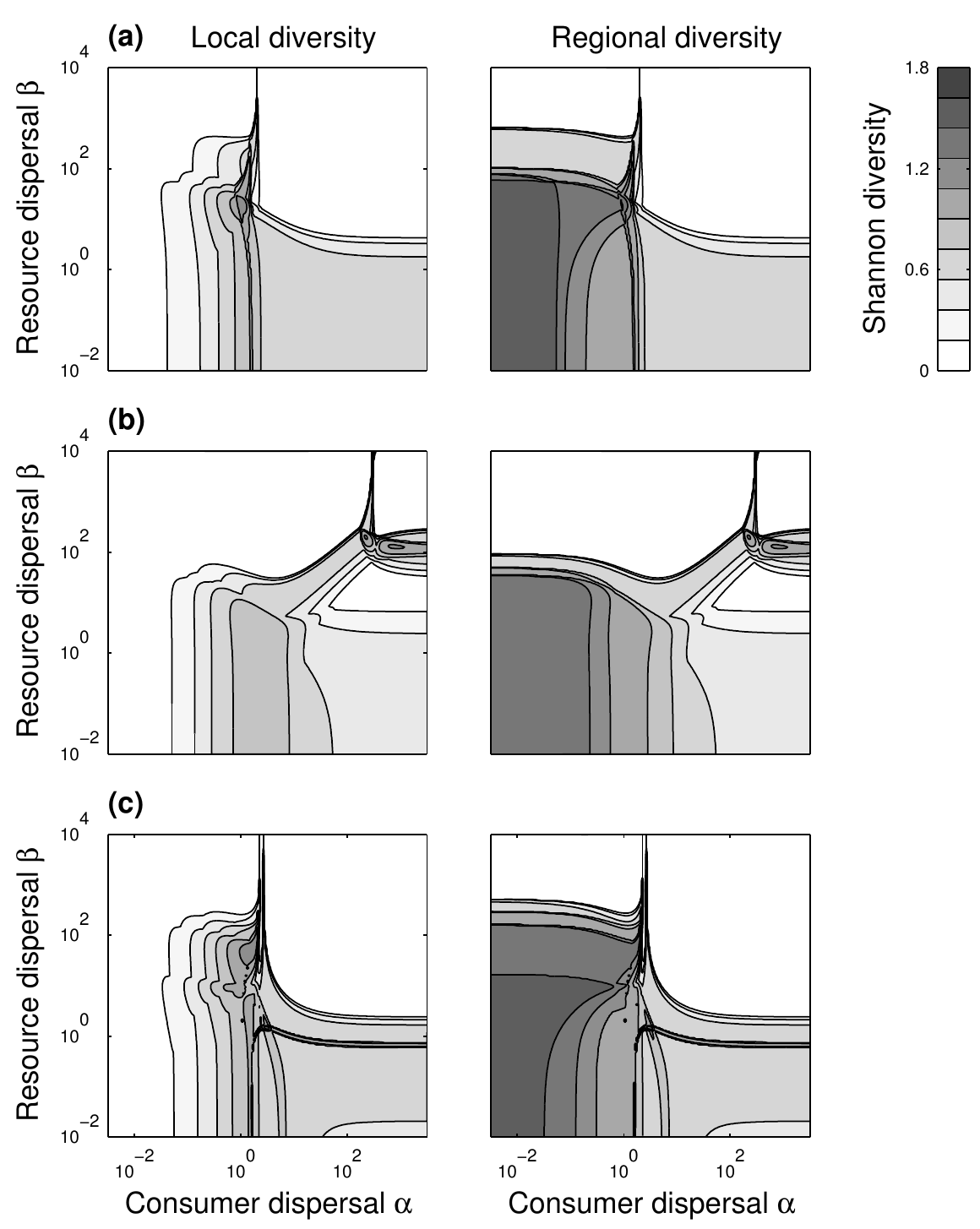}
\caption{Diversity patterns for five-patch metacommunities without and with trade-offs. Local and regional diversity are plotted as a function of consumer dispersal $\alpha$ and resource dispersal $\beta$ for three metacommunities with five patches and a species pool consisting of 20 species.  Panel (a):  Consumption rates are drawn independently from an exponential distribution with mean 1.  Panel (b):  Consumption rates are drawn randomly with a linear trade-off given by $\sum_k c_{ik} = 5 \pm 0.05$.  Panel (c):  Consumption rates are drawn randomly with a quadratic trade-off given by $\sum_k c_{ik}^2 = 9 \pm 0.1$.  Abiotic resource with $a=1$ and patch fertilities $A_k$ drawn independently from an exponential distribution with mean 2.  Other parameters: $e = m = 1$.}
\end{center}
\end{figure}

\clearpage\newpage

%%% FIGURE S9 %%%
\begin{figure}
\begin{center}
\includegraphics[width=.84\textwidth]{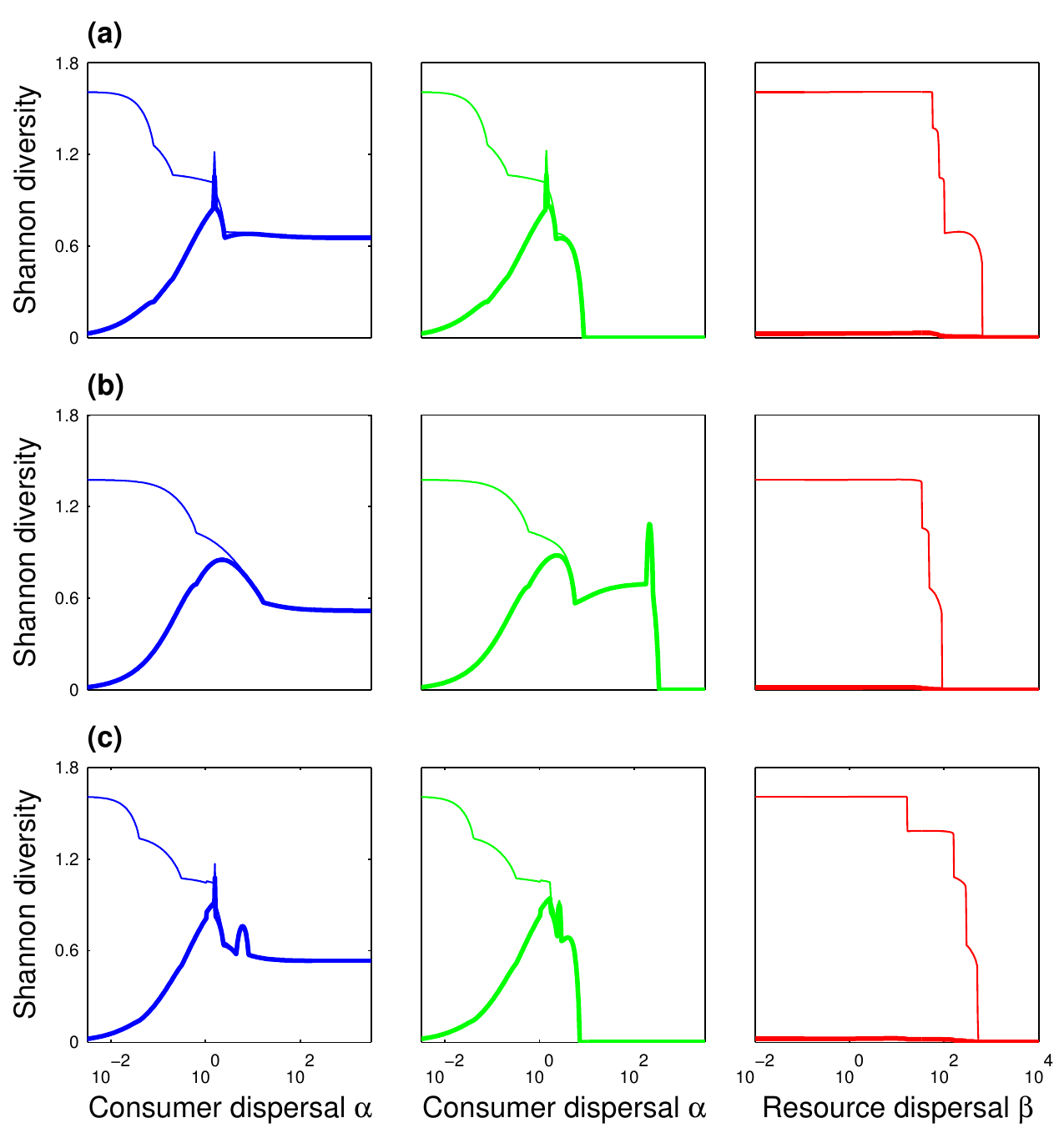}
\caption{One-dimensional diversity-dispersal relationships for five-patch metacommunities without and with trade-offs.  Local diversity (thick line) and regional diversity (thin line) are plotted as a function of consumer dispersal $\alpha$ or resource dispersal $\beta$.  Same metacommunities as in Figure~S8.  Same color code as in Figure~4, that is, blue: variable $\alpha$ and constant $\beta$;  green: $\alpha$ and $\beta$ change simultaneously;  red: constant $\alpha$ and variable $\beta$.}
\end{center}
\end{figure}

\clearpage\newpage

%%% FIGURE S10 %%%
\begin{figure}
\begin{center}
\includegraphics[width=.76\textwidth]{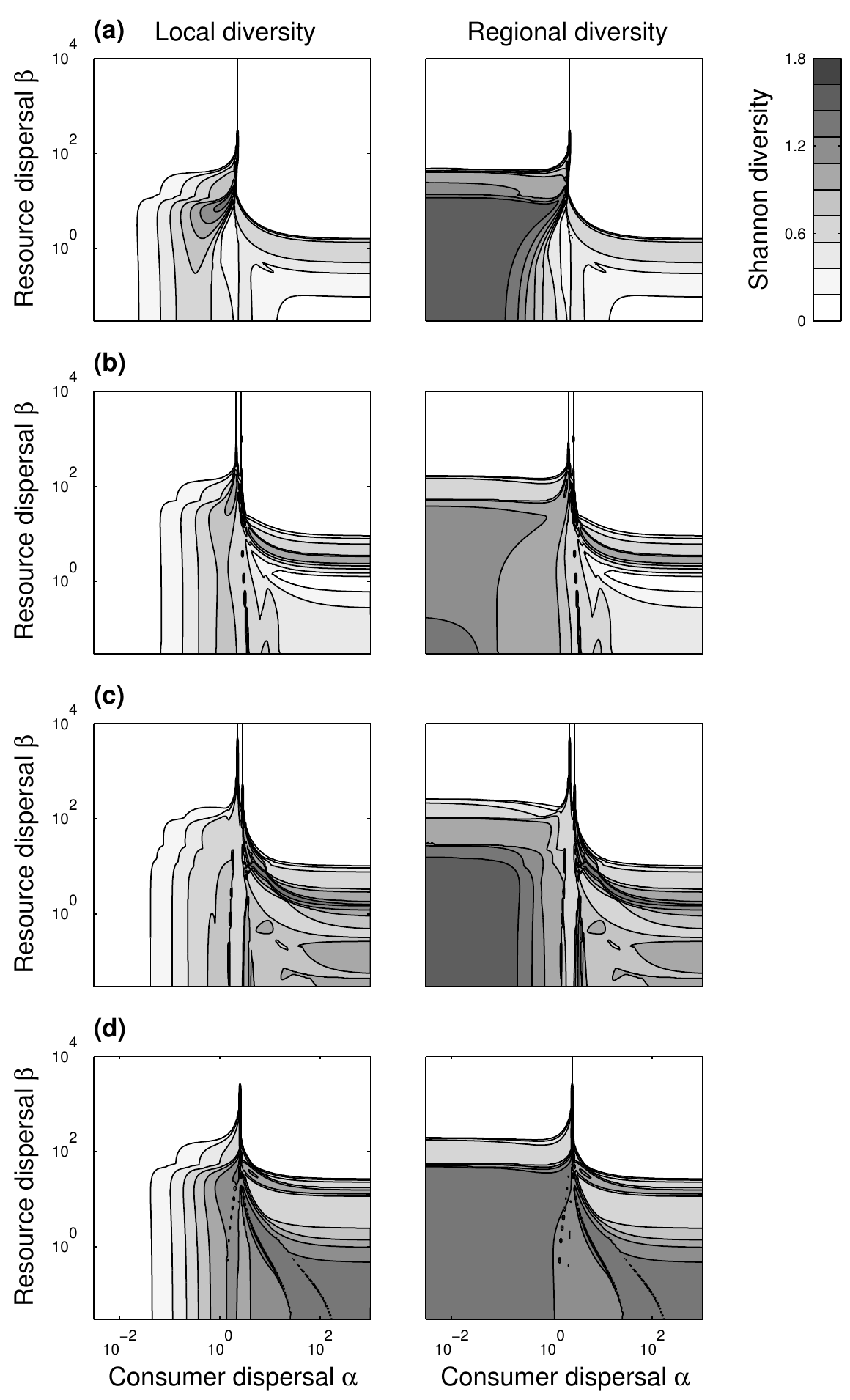}
\caption{Diversity patterns for five-patch metacommunities with quadratic trade-off.  Local and regional diversity
are plotted as a function of consumer dispersal~$\alpha$ and resource dispersal~$\beta$ for four metacommunities with five patches and a species pool consisting of 20 species.  Consumption rates are drawn randomly with a quadratic trade-off given by $\sum_k c_{ik}^2 = 25 \pm 0.5$.  Biotic resource with $b=1$ and patch fertilities $B_k$ drawn independently from an exponential distribution with mean 2.  Other parameters: $e = m = 1$.}
\end{center}
\end{figure}

\clearpage\newpage

%%% FIGURE S11 %%%
\begin{figure}
\begin{center}
\includegraphics[width=.84\textwidth]{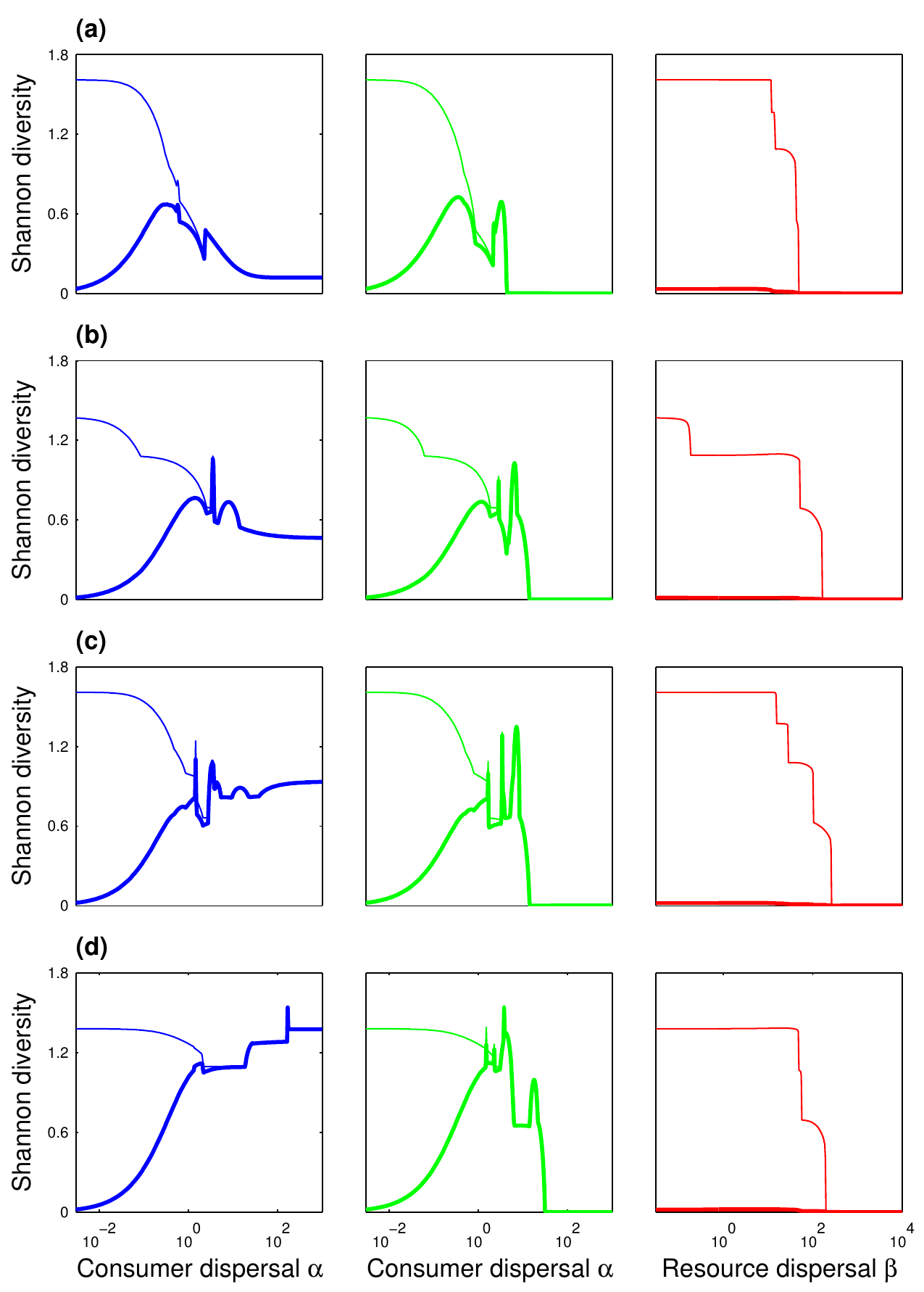}
\caption{One-dimensional diversity-dispersal relationships for five-patch metacommunities with quadratic trade-off.  Local diversity (thick line) and regional diversity (thin line) are plotted as a function of consumer dispersal $\alpha$ or resource dispersal $\beta$.  Same metacommunities as in Figure~S10.  Same color code as in Figure~4, that is, blue: variable $\alpha$ and constant $\beta$;  green: $\alpha$ and $\beta$ change simultaneously;  red: constant $\alpha$ and variable $\beta$.}
\end{center}
\end{figure}

\clearpage\newpage

%%% FIGURE S12 %%%
\begin{figure}
\begin{center}
\includegraphics[width=.58\textwidth]{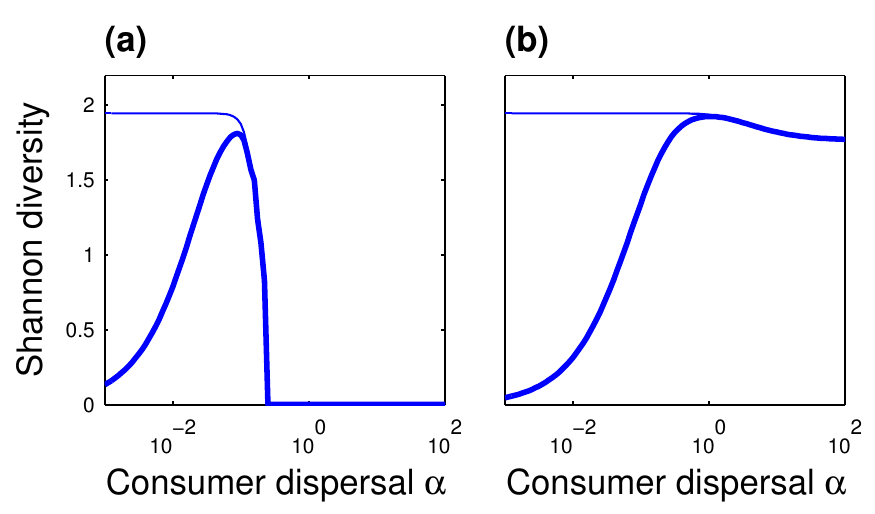}
\caption{Diversity-dispersal relationships predicted by Loreau et al.~(2003a).  Local diversity (thick line) and regional diversity (thin line) are plotted as a function of consumer dispersal $\alpha$ with resource dispersal $\beta = 0$.  Panel (a):  Parameter values of Loreau et al.~(2003a) with consumption rates (\ref{eq:consrate1}).  Panel (b):  Modified parameter values with consumption rates (\ref{eq:consrate2}).}
\end{center}
\end{figure}

\end{document}